\documentclass[aip,numerical,jcp,reprint,amsmath,amssymb]{revtex4-1}

\usepackage[font=small, labelfont=bf, justification=justified,singlelinecheck=false]{caption}
\usepackage{enumerate}

\usepackage{amsmath}
\usepackage{amssymb}

\usepackage{cancel}
 \usepackage{caption}
 \usepackage{subcaption}
\usepackage{ifpdf}
\usepackage{rotating}
\usepackage{alltt}
\usepackage{units}
\usepackage{color}
\usepackage[usenames,dvipsnames]{xcolor}

\definecolor{lgray}{gray}{0.7}
\definecolor{llgray}{gray}{0.5}
\definecolor{lllgray}{gray}{0.3}

\usepackage{pstricks}
\usepackage{pst-node}
\usepackage{pst-blur}
\definecolor{Pink}{rgb}{1.,0.75,0.8}

\usepackage{mdwlist}

\usepackage{setspace}

\newcommand*{\xlineshort}[1][1.2em]{\rule[0.4ex]{3.5pt}{0.5pt}}

\newcommand*{\xdash}[1][1.2em]{\rule[0.4ex]{2.5pt}{0.5pt} \rule[0.4ex]{2.5pt}{0.5pt}}
\newcommand*{\xdashthick}[1][1.2em]{\rule[0.4ex]{3.5pt}{1.5pt} \rule[0.4ex]{3.5pt}{1.5pt}}
\newcommand*{\xdashvthick}[1][1.2em]{\rule[0.4ex]{4.5pt}{2.5pt} \rule[0.4ex]{4.5pt}{2.5pt}}

\newcommand{\eq}[1]{ Eq.\ (\ref{#1})}

\DeclareMathOperator*{\Pressuretype}{\textbf{P}}

\DeclareMathOperator*{\heatflux}{{\bf{J}}_{\it{q}}}

\newcommand{\ie}{{i.e.\ }}

\newcommand{\tens}[1]{\boldsymbol{#1}}

\DeclareMathOperator*{\define}{\equiv}

\DeclareMathOperator*{\dV}{\it{d}\textbf{r}}
\DeclareMathOperator*{\intV}{\Delta {\it{V}}}

\newcommand{\isum}{ \displaystyle\sum_{i=1}^N}

\newcommand{\ijsum}{\displaystyle\sum_{i,j}^N}

\DeclareMathOperator*{\MDvel}{\textbf{v}_{\mathnormal{i}}}
\DeclareMathOperator*{\MDpvel}{{\textbf{c}}_{\mathnormal{i}}}
\DeclareMathOperator*{\MDaccel}{\ddot{{\bf{r}}}_{\mathnormal{i}}}
\DeclareMathOperator*{\MDenergy}{\mathnormal{e_i}}
\DeclareMathOperator*{\MDpenergy}{\mathnormal{u_i}}

\DeclareMathOperator*{\CFDvel}{\textbf{v}}

\DeclareMathOperator*{\Fij}{{\bf{F}}_{\it{{ij}}}}
\DeclareMathOperator*{\Fijrij}{{\bf{r}}_{\it{{ij}}} \Fij}

\begin{document}
\title{Measuring Heat Flux Beyond Fourier's law}

\author{E. R. Smith}
\affiliation{Department of Mechanical Engineering, Imperial College London, Exhibition Road, South Kensington, London SW7 2AZ, United Kingdom} 

\author{P. J. Daivis}
\affiliation{School of Science and Centre for Molecular and Nanoscale Physics, RMIT University, GPO Box 2476, Melbourne, Victoria, 3001, Australia} 

\author{B. D. Todd}
\affiliation{Department of Mathematics, Faculty of Science, Engineering and Technology, Swinburne University of Technology, P.O. Box 218, Hawthorn, Victoria, 3112, Australia}

\date{\today}

\begin{abstract}
\section*{Abstract}

We use nonequilibrium molecular dynamics (NEMD) to explore the effect of shear flow on heat flux. 
By simulating a simple fluid in a channel bounded by tethered atoms, the heat flux is computed for two systems: a temperature driven one with no flow and a wall driven, Couette flow system. The results for the temperature driven system give the Fourier's law thermal conductivity, which is shown to agree well with experiments.
Through comparison of the two systems, we quantify the additional components of the heat flux parallel and normal to the walls due to shear flow. 
To compute the heat flux in the flow direction, the Irving-Kirkwood equations are integrated over a volume, giving the so-called volume average form, and they are also manipulated to get expressions for the surface averaged and method of planes forms.
The method of planes and volume average forms are shown to give equivalent results for the heat flux when using small volumes.
The heat flux in the flow direction is obtained consistently over a range of simulations, and it is shown to vary linearly with strain rate, as predicted by theory.
The additional strain rate dependent component of the heat flux normal to the wall is obtained by fitting the strain rate dependence of the heat flux to the expected form.
As a result, the additional terms in the thermal conductivity tensor quantified in this work should be experimentally testable. 

\end{abstract}

\maketitle

\section{Introduction}
\label{sec:Intro}

In his 1878 work, Joseph Fourier proposed the existence of a single coefficient of thermal conductivity to describe heat conduction in an infinite solid \citep{Freeman_Fourier_1878}.
The simple linear relationship between heat flux and temperature gradient is known as Fourier's law, an empirical observation of great importance in many areas of science and engineering.
However, there are cases when a single coefficient of thermal conductivity is not sufficient to model the full range of physics. 
Examples include systems under extremely high shear rates as observed in tribological applications, nanoscopic systems where interfacial effects dominate and highly inhomogeneous systems where local constitutive equations break down.
In these cases, a detailed understanding of heat flux at the small scale is essential. Deviations from Fourier's law are potentially significant in nanochannels, MEMS and NEMS, nanoscale cpu components and complex materials such as aerogels with nano-scale pores, but they are exceptionally difficult to measure. It is in this context that molecular dynamics (MD) simulations are uniquely placed to provide insight.

Reliable methods for obtaining the Fourier's law thermal conductivity are well known, including the Green-Kubo \citep{Green, Kubo}, Evans heat flow algorithm \citep{EVANS1982457} and sinusoidal transverse field (STF) methods \citep{Gosling_et_al}. The thermal conductivity can also be obtained by comparing temperature profiles obtained in inhomogeneous nonequilibrium molecular dynamics (NEMD) simulations with predictions of constitutive laws \citep{todd_daivis_2017}.

When the thermodynamic forces are large, the fluxes may depend on them nonlinearly. Under these conditions, Curie's principle no longer applies and the heat flux may depend on nonlinearly coupled combinations of the temperature gradient and the velocity gradient \citep{daivis1993generalized}. If the velocity gradient varies rapidly with position, nonlocal effects also become important and the second spatial derivative of the velocity field may also be required as a thermodynamic force \citep{Baranyai_Evans_Daivis_1992, daivis2000generalized}. Both of these effects lead to a component of the heat flux vector parallel to the streaming velocity in addition to a modification of the usual perpendicular heat flow due to Fourier's Law. These results can be summarised in a constitutive equation of the form given by Daivis and Coelho \citep{daivis2000generalized}. For a system in steady planar shear with a velocity gradient of the form $\nabla {\mathbf{v}} = \dot \gamma \left( y \right){\mathbf{ji}}$, a truncated 
Taylor expansion of the heat flux vector in powers of the thermodynamic forces $\nabla {\mathbf{v}}$, $\nabla \nabla{\mathbf{v}}$ and $\nabla T$ gives

\begin{equation}
\label{Jq_constit}
{{\mathbf{J}}_q}^{approx} =  - {\tens{\lambda_{\textrm{eff}}}} \cdot \nabla T + \zeta \frac{{\partial \dot \gamma }}{{\partial y}}{\mathbf{i}} + 2\xi \dot \gamma \frac{{\partial \dot \gamma }}{{\partial y}}{\mathbf{j}} .
\end{equation}

In a flow with a uniform velocity gradient field, such as homogeneous planar shear flow, ${\partial \dot \gamma }/{\partial y}$ is zero and only the first term is relevant.

The thermal conductivity tensor for a homogeneously shearing fluid with the fluid velocity in the $x$ direction and the gradient in the $y$ direction, i.e. with $\nabla {\mathbf{v}} = \dot \gamma \left( y \right){\mathbf{ji}}$, is given by
\begin{equation}
\label{tc_tensor}
{\tens{\lambda_{\textrm{eff}}}} = \left[ {\begin{array}{*{20}{c}}
  {\lambda  + 3{\lambda_{2}}{{\dot \gamma }^2}}&{-{\lambda_1}\dot \gamma }&0 \\ 
  {-{\lambda_1}\dot \gamma }&{\lambda + 3{\lambda_{2}}{{\dot \gamma }^2}}&0 \\ 
  0&0&{\lambda  + {\lambda_{2}}{{\dot \gamma }^2}} 
\end{array}} \right] .
\end{equation}
The three thermal conductivity coefficients are: $\lambda$, the linear Fourier's law thermal conductivity, $\lambda_2$ the coefficient of a nonlinear shear rate dependent heat flux in the direction of the temperature gradient ($y$) and $\lambda_1$ the coefficient for the shear induced heat flux in the flow direction, parallel to the walls ($x$).

Note that the off diagonal terms are chosen to be negative to give positive values of $\lambda_1$, a different convention than used previously \citep{daivis1993generalized}. The linear dependence on shear rate for the off-diagonal and quadratic dependence for the diagonal components of the thermal conductivity tensor has been confirmed for a model of liquid butane by \citet{Daivis_Evans1995}.

The tensor character of the generalised thermal conductivity reflects the anisotropy of the fluid induced by the shearing velocity gradient field. 
While shear rate dependent contributions to the heat flux should in principle exist, it has been suggested that it would be very challenging to measure them experimentally \citep{Todd_Evans_95}. In fact very few measurements of anisotropic thermal conductivity due to shear flow exist \cite{Venerus2004}.

Kinetic theory, using Bhatnagar-Gross-Krook and Grad 13-moment style schemes has also confirmed the tensor character of the thermal conductivity for a fluid in planar shear flow \citep{Garzo_1994}, as well as the existence of a secondary heat flux parallel to the streamlines, driven purely by the gradient of the shear rate, in dilute gases under planar Poiseuille flow \citep{risso1997dilute}. Using the STF method, the shear-rate gradient induced heat flux was shown by \citet{Baranyai_Evans_Daivis_1992} to result in as much as a 25\% departure from the purely linear solution for the temperature profile. Heat flux in Poiseuille flow of a dense fluid was also obtained from NEMD methods by \citet{todd1997temperature}. The magnitude of the second coefficient was shown in shear roller flow as a function of wavenumber \citep{Monaghan_Morriss_97} for 2D flow. \citet{Han_Lee} also proposed a local form of the method of planes formalism \cite{Todd_et_al_95, Todd_Evans_Daivis_95, Daivis_et_al} and captured the parallel 
component of the heat flux 
in an NEMD simulation. The strain rate coupling was also explored by \citet{Menzel_et_al} using NEMD for a polymer solution which indicates a very small value of the strain rate coupling coefficient $\xi$ with a magnitude of approximately one in reduced units, in agreement with literature values.

In this work, we explore the heat flux in planar shear flow using inhomogeneous NEMD simulations. In order to accurately compute the contribution of the parallel component, we use a large molecular system with approximately a quarter of a million atoms over a million timesteps. To isolate the shear rate dependent contribution to the heat flux from other contributions, we compare two systems; one which applies only a temperature gradient by maintaining different temperatures at the two channel walls with no strain rate and a second system with planar shear flow which generates its own temperature gradient and equal temperatures at the thermostatted walls. By running a parameter study over a range of densities in both systems, we isolate the shear rate dependent components of the thermal conductivity as a function of shear rate at each density.

We consider wall driven planar shear flow and obtain the mass density, streaming velocity, pressure tensor and heat flux vector using the formulation of \citet{Irving_Kirkwood}, integrated over either a cuboidal volume in space or a small rectangular surface. Spatial integration or averaging is important as it allows terms involving the Dirac delta functions in the Irving-Kirkwood equations to be numerically computed \citep{Smith_et_al}. The fluxes can then be written in one of two forms; $i)$ the volume average (VA) form \citep{Cormier_et_al, Hardy, Lutsko} and $ii)$ a surface averaged form, which is a generalisation of the method of planes (MOP) \citep{Todd_et_al_95, Todd_Evans_Daivis_95, Daivis_et_al, Heinz_et_al05} providing \textit{all} components of the pressure tensor and heat flux vector on the localised surfaces of a volume \citep{Han_Lee}.

This paper is organised as follows: in Section \ref{sec:Theory}, the formulae for obtaining the heat flux are derived in both volume averaged and surface averaged forms. Next, the extended constitutive equations relevant to the current systems are outlined. In Section \ref{sec:Simulation} the details of the two NEMD simulations, conducted on systems 1 and 2, are outlined. The results and discussion are presented next in Section \ref{sec:Results}. Finally, this work finishes with a summary and conclusions in Section \ref{sec:Conclusions}.
\vspace{-0.2in}
\section{Theory}
\label{sec:Theory}

Expressions for the pressure tensor and heat flux vector can be derived by following the \citet{Irving_Kirkwood} procedure. They typically begin \citep{Evans_Morris} with definitions of the local instantaneous mass density $\rho$, momentum density $\rho {\mathbf{v}}$ and total energy density $\rho e$,
\begin{align}
  \rho \left( {{\mathbf{r}},t} \right) &= \sum\limits_{i = 1}^N {{m_i}\delta \left( {{\mathbf{r}} - {{\mathbf{r}}_i}} \right)}   \\
  \rho {\mathbf{v}}\left( {{\mathbf{r}},t} \right) &= \sum\limits_{i = 1}^N {{m_i}{{\mathbf{v}}_i}\delta \left( {{\mathbf{r}} - {{\mathbf{r}}_i}} \right)}  \\
  \rho e\left( {{\mathbf{r}},t} \right) &= \sum\limits_{i = 1}^N {{e_i}\delta \left( {{\mathbf{r}} - {{\mathbf{r}}_i}} \right)}
\end{align}
where ${m_i}$, ${{\mathbf{v}}_i}$ and ${e_i}$ are the mass, velocity and total energy of atom $i$. The delta function (strictly speaking a generalised function) represents the atomic localisation of each property.
The densities are substituted into the momentum and energy balance equations,
\begin{align}
  \frac{{\partial \left( {\rho {\mathbf{v}}} \right)}}{{\partial t}} &=  - \nabla  \cdot \left( {\mathbf{P} + \rho {\mathbf{vv}}} \right) \hfill \\
  \frac{{\partial \left( {\rho e} \right)}}{{\partial t}} &=  - \nabla  \cdot \left( {{{\mathbf{J}}_q} + \rho e{\mathbf{v}} + \mathbf{P} \cdot {\mathbf{v}}} \right),
\end{align}
leading to explicit expressions \citep{Evans_Morris,todd_daivis_2017} for the local instantaneous pressure tensor $\mathbf{P}\left( {{\mathbf{r}},t} \right)$ and heat flux vector ${{\mathbf{J}}_q}\left( {{\mathbf{r}},t} \right)$,
\begin{align}
\label{ptlocal}
  \mathbf{P}\left( {{\mathbf{r}},t} \right) = \sum\limits_i^N {{m_i}{{\mathbf{c}}_i}{{\mathbf{c}}_i}\delta \left( {{\mathbf{r}} - {{\mathbf{r}}_i}} \right)} \;\;\;\;\;\;\;\;\;\;\;\;\;\;\;\;\;\;\;\;\;\;\;\;\;\;\;\;\;\;\;\;\;\;\;\;\;\;\;
  \nonumber \\
  - \frac{1}{2}\sum\limits_i^N {\sum\limits_{j \ne i}^N {{{\mathbf{r}}_{ij}}{{\mathbf{F}}_{ij}}\int_0^1 {\delta \left( {{\mathbf{r}} - {{\mathbf{r}}_i} - s {{\mathbf{r}}_{ij}}} \right)ds } } }   \\
\label{jqlocal}
  {{\mathbf{J}}_q}\left( {{\mathbf{r}},t} \right) = \sum\limits_i^N {{u_i}{{\mathbf{c}}_i}\delta \left( {{\mathbf{r}} - {{\mathbf{r}}_i}} \right)} \;\;\;\;\;\;\;\;\;\;\;\;\;\;\;\;\;\;\;\;\;\;\;\;\;\;\;\;\;\;\;\;\;\;\;\;\;\;\;
  \nonumber \\
  - \frac{1}{2}\sum\limits_i^N {\sum\limits_{j \ne i}^N {{{\mathbf{r}}_{ij}}{{\mathbf{F}}_{ij}} \cdot \left[ {{{\mathbf{v}}_i} - {\mathbf{v}}\left( {{\mathbf{r}},t} \right)} \right]\int_0^1 {\delta \left( {{\mathbf{r}} - {{\mathbf{r}}_i} - s {{\mathbf{r}}_{ij}}} \right)ds } } } 
\end{align}
where $u_i $ is the internal energy of particle $i$, ${\mathbf{c}}_i = {{{\mathbf{v}}_i} - {\mathbf{v}}\left( {{\mathbf{r}_i},t} \right)}$ is its thermal or peculiar velocity and ${\mathbf{r}}_{ij} = {\mathbf{r}}_{j} - {\mathbf{r}}_{i}$.
The Irving-Kirkwood expressions for the local pressure tensor and heat flux vector are well established, but they are purely formal. They contain delta functions which must be integrated in order to generate numerical values for the fluxes. In this section, volume averaged and surface averaged expressions for the heat flux vector are derived by integrating the \citet{Irving_Kirkwood} fluxes over either the volume or the surfaces of a three-dimensional cuboid in Cartesian coordinates. When this cuboid is sufficiently small we obtain good approximations to the local values of the pressure tensor and heat flux vector. This presentation is kept brief as the majority has been given previously \citep{Smith_et_al, Heyes_et_al14}.

\subsection{Volume average fluxes}
\label{sec:VA_derivation}
 To obtain the volume averaged form of the pressure tensor and heat flux vector, we integrate both sides of the flux expressions in space. The integration volume $\Delta V$ can be regarded as a control volume in the fluid dynamics sense \citep{Hirsch, Potter_Wiggert}. Integrating the densities and dividing by the volume gives the average densities
\begin{align}
\rho_V &= \frac{1} {\Delta V} \int_{\intV} \rho \dV = \frac{1} {\Delta V} \displaystyle\sum_{i=1}^N   m_i \vartheta_i \\
\left({\rho \CFDvel}\right)_V &= \frac{1} {\Delta V}\int_{\intV} \rho \CFDvel \dV = \frac{1} {\Delta V} \displaystyle\sum_{i=1}^N  m_i  \MDvel \vartheta_i \\ 
\left(\rho e\right)_V &= \frac{1} {\Delta V} \int_{\intV}  \rho e \dV = \frac{1} {\Delta V} \displaystyle\sum_{i=1}^{N} \MDenergy \vartheta_i
\end{align}
within volume $\Delta V$. Integrating the delta function gives a value of $1$ for each particle within $\intV$ and zero for each particle outside that volume \citep{Smith_et_al}. This is represented by the spatial selector function $\vartheta _i$ which can be written as
\begin{align}
  {\vartheta _i} = \int_{\Delta V} {\delta \left( {{\mathbf{r}} - {{\mathbf{r}}_i}\left( t \right)} \right)d{\mathbf{r}}}  
  \nonumber \\
   = \int_{-\infty}^{\infty} {\vartheta \left( {\frac{{{\mathbf{r}} - {{\mathbf{r}}_m}}}{{\Delta {\mathbf{r}}}}} \right)\delta \left( {{\mathbf{r}} - {{\mathbf{r}}_i}\left( t \right)} \right)d{\mathbf{r}}}  
   = \vartheta \left( {\frac{{{{\mathbf{r}}_i} - {{\mathbf{r}}_m}}}{{\Delta {\mathbf{r}}}}} \right) 
\end{align}

The $\vartheta$ function is a three dimensional generalisation of the rectangle function defined by \citet{Bracewell},
\begin{equation}
\label{thetadef}
\vartheta \left( {\frac{{{\mathbf{r}} - {{\mathbf{r}}_m}}}{{\Delta {\mathbf{r}}}}} \right) = 
\nonumber \\
\Pi \left( {\frac{{x - {x_m}}}{{\Delta x}}} \right)\Pi \left( {\frac{{y - {y_m}}}{{\Delta y}}} \right)\Pi \left( {\frac{{z - {z_m}}}{{\Delta z}}} \right)
\end{equation}
where the one dimensional rectangle function is defined by
\begin{equation}
\label{rectdef}
\Pi \left( {\frac{{x - {x_m}}}{{\Delta x}}} \right) = \left[ {H\left( {x - {x_ - }} \right) - H\left( {x - {x_ + }} \right)} \right]
\end{equation}
and $H\left( {x} \right)$ is the Heaviside step function. Similar relations hold for the rectangle functions in the $y$ and $z$ directions. In the definition of the rectangle function, ${x_m}$ is the midpoint of the rectangle $x_m = (x_- - x_+)/2$, and ${\Delta x}$ is its width, and it is assumed that ${x_ + }>{x_ - }$.

Although we are mainly concerned with the heat flux vector in this paper, it is worth considering the simpler case of the pressure tensor first.
Because the kinetic part of the pressure tensor is essentially a density, its volume average is easily evaluated in this form,
\begin{align}
\int_{\intV}  \Pressuretype{}^K (\textbf{r},t) \dV = \displaystyle\sum_{i=1}^{N}  m_i \MDpvel \MDpvel \vartheta_i
\end{align}
where $\MDpvel$ is the peculiar velocity $\MDpvel \define \MDvel - \CFDvel(\textbf{r}_i,t)$, and the average velocity $\CFDvel$ can be evaluated from the locally averaged momentum and mass densities
\begin{align}
{\mathbf{\bar v}}\left( {{{\mathbf{r}}_m},t} \right) \define \frac{\int_{\intV}  \rho \CFDvel \dV}{\int_{\intV}  \rho \dV} = \frac{\sum_{i=1}^N  m_i \MDvel \vartheta_i}{\sum_{i=1}^N  m_i \vartheta_i}
\label{vel_definition}
\end{align}
and $\textbf{v}_i$ is the total velocity of particle $i$, i.e. the sum of the thermal and streaming velocities, with the streaming velocity at the location of the particle $\textbf{r}_i$ while the velocity obtained from \eq{vel_definition} is the average for a volume with centre at ${\mathbf{r}}_m$.

Next, let us consider the configurational part of the pressure tensor,
\begin{align}
\int_{\intV}  \Pressuretype{}^\phi (\textbf{r},t) \dV = - \frac{1}{2}   \ijsum  \Fijrij  \int_{\intV} \int_0^1  \delta(\bf{r} - \bf{r}_s) ds  \dV , 
 \label{pressure_configure}
\end{align}
where ${{\bf{r}}_\lambda } = {{\bf{r}}_i} + s {{\bf{r}}_{ij}}$ is a point on the line between molecules $i$ and $j$ and we use the subscript $i,j$ on the summation to denote sum over all $i$ and all $j$ with $i \ne j$. 
Swapping the order of integration, inserting the three dimensional rectangle function to limit the range of integration and using the sifting property of the delta function, we obtain
\begin{align}
  \int_{\Delta V}^{} {\int_0^1 {\delta \left( {{\mathbf{r}} - {{\mathbf{r}}_s }} \right)ds } } d{\mathbf{r}} 
  \nonumber \\  
  = \int_0^1 {\int_{ - \infty }^\infty  {\vartheta \left( {\frac{{{\mathbf{r}} - {{\mathbf{r}}_m}}}{{\Delta {\mathbf{r}}}}} \right)\delta \left( {{\mathbf{r}} - {{\mathbf{r}}_s }} \right)d{\mathbf{r}}ds } }  
   \nonumber \\  
  = \int_0^1 {{\vartheta \left( {\frac{{{\mathbf{r}_s} - {{\mathbf{r}}_m}}}{{\Delta {\mathbf{r}}}}} \right) ds } }.
\end{align}
The integrand on the right hand side is denoted by $\vartheta_s$, a function which is one if the point ${\bf{r}}_s$ is inside $\Delta V$ and zero otherwise, with the whole right hand side typically expressed as,
\begin{align}
 \ell_{ij} \define \int_0^1 \vartheta_s ds,
 \label{int_s}
\end{align}
where the integral is equal to the fraction of the line between particles $i$ and $j$ that is enclosed by the volume $\intV$. 
The average of the local pressure over a small volume $\intV$ is $\Pressuretype_V = \left( 1/ \intV \right) \int_{\intV}  \Pressuretype (\textbf{r},t) \dV$. Inserting the microscopic expressions, the volume averaged pressure tensor is then,
\begin{align}
\Pressuretype{}_V (\textbf{r},t) =\frac{1}{{\intV} } \left[  \displaystyle\sum_{i=1}^{N}  m_i \MDpvel \MDpvel \vartheta_i
 -  \frac{1}{2}   \displaystyle\sum_{i,j}^{N} \Fijrij  \ell_{ij} \right]. 
 \label{pressure}
\end{align}
To obtain the volume averaged form of the heat flux vector, we follow a similar procedure.
Similarly to the pressure tensor, the average of the kinetic part of the local heat flux vector \eq{jqlocal} over a small volume $\intV$ is,
\begin{align}
   \heatflux{\!\!\!}^K  = \frac{1}{\intV}\displaystyle\sum_{i=1}^N  \MDpenergy \MDpvel \vartheta_i.
\end{align}
To compute the volume average of the configurational part of \eq{jqlocal}, we change the order of integration, introduce the three dimensional rectangle function and use the properties of the delta function to find
\begin{align}
\int_{\intV} \CFDvel (\textbf{r},t) \int_0^1  \delta(\textbf{r} - \textbf{r}_s) d s  \dV 
\nonumber \\ 
= \int_0^1  \CFDvel (\textbf{r}_s, t) \vartheta_s d s \approx {\mathbf{\bar v}}\left( {{{\mathbf{r}}_m},t} \right).
\end{align}
where ${\mathbf{\bar v}}\left( {{{\mathbf{r}}_m}}, t \right)$ is the volume averaged velocity at $\mathbf{r}_m$, discussed in the appendix section \ref{sec:A_derivation}. 
This assigns the velocity contributions based on interaction length inside a volume, as discussed further in the appendix.
Instead of evaluating the velocity integrated along the line, the velocity is approximated as the average value in the same volume in which we have evaluated the pressure.
For simplicity, as we are using volume averages for other quantities, we assume the fluid velocity is also constant inside $\Delta V$, so we can write
\begin{align}
  \heatflux{}_V  =  \frac{1}{\intV}\left[ \displaystyle\sum_{i=1}^N  \MDpenergy \MDpvel \vartheta_i -  \frac{1}{2} \displaystyle\sum_{i,j}^N \Fijrij \cdot \left( \MDvel  - {\mathbf{\bar v}}\left( {{{\mathbf{r}}_m},t} \right) \right) \ell_{ij} \right] . 
\label{EnergyEqminusadvct2_VA}
\end{align}
Error will be introduced into the computation of the heat flux vector if this relation is used when $\CFDvel$ is not constant, but the error can be minimised by reducing the dimensions of $\Delta V$.

For practical computation it is convenient to split the heat flux into kinetic and configurational parts, so the total volume average heat flux is $\int_{\intV} \heatflux \dV = \int_{\intV} \left( \heatflux^{\!\!\!K} + \heatflux^{\!\!\! \phi} \right) \dV$, that is, using the sum of the volume averaged kinetic and configurational components. 
Then the volume averaged form of the kinetic part of the heat flux vector suitable for computation is
\begin{equation}
{\mathbf{J}}_{qV}^K\left( {{{\mathbf{r}}_m},t} \right) = \frac{1}{{\Delta V}}\left[ {\sum\limits_{i = 1}^N {{e_i}{{\mathbf{v}}_i}{\vartheta _i}}  - {\mathbf{\bar v}}\left( {{{\mathbf{r}}_m},t} \right)\sum\limits_{i = 1}^N {{e_i}{\vartheta _i}} } \right] .
   \label{VA_advection}
\end{equation}

If the streaming velocity is approximately constant within $\Delta V$, the configurational component of the heat flux vector should be well approximated by
\begin{align}
{\mathbf{J}}_{qV}^\phi \left( {{{\mathbf{r}}_m},t} \right) = \;\;\;\;\;\;\;\;\;\;\;\;\;\;\;\;\;\;\;\;\;\;\;\;\;\;\;\;\;\;\;\;\;\;\;\;\;\;\;\;\;\;\;\;\;\;\;\;\;\;\;\;\;\;\;\;\;\;\;\;\;\;\;\;\;\;\;\;\;\;\;
\nonumber \\
 - \frac{1}{{\Delta V}}\frac{1}{2}\left[\ijsum {{{\mathbf{r}}_{ij}}{{\mathbf{F}}_{ij}} \cdot {{\mathbf{v}}_i}{\ell _{ij}} 
- \left( \ijsum {{{\mathbf{r}}_{ij}}{{\mathbf{F}}_{ij}}{\ell _{ij}}} \right) \cdot {\mathbf{\bar v}}\left( {{{\mathbf{r}}_m}} \right)} \right].
   \label{VA_stressheat}
\end{align}

In this form, the average velocities, average pressure and other quantities can be accumulated during the simulation until satisfactory statistics are achieved and then both components of the heat flux vector can be calculated at the end of the run.

\subsection{Surface averaged fluxes}
\label{sec:MOP_derivation}
When the fluxes are averaged over a finite surface, we obtain generalisations of the method of planes results of \citet{Todd_et_al_95, Todd_Evans_Daivis_95} as given in \citet{Han_Lee} and \citet{Heinz_et_al05}. 
A mathematical expression for fluxes over a volume surface in terms of functionals was derived previously for the pressure tensor \citep{Smith_et_al} and is extended here for the heat flux vector.

The integral of a tensor over a closed surface is related to the volume integral of its divergence through the divergence theorem. For the pressure tensor, which is a second rank tensor, this is
\begin{equation}
\oint_A {{\mathbf{P}^T} \cdot d{\mathbf{A}}}  = \int_{\Delta V} {\nabla  \cdot \mathbf{P}d{\mathbf{r}}} .
\end{equation}

This relationship gives us a convenient way to obtain the surface averaged fluxes from the local expressions for the pressure tensor, \eq{ptlocal}, and heat flux vector, \eq{jqlocal}. Applying the divergence theorem to the kinetic part of the pressure tensor, we find
\begin{align}
  \int_{\Delta V} {\nabla  \cdot {\mathbf{P}^K}d{\mathbf{r}}}  &= \isum {{m_i}{{\mathbf{c}}_i}{{\mathbf{c}}_i} \cdot \int_{\Delta V} {\nabla \delta \left( {{\mathbf{r}} - {{\mathbf{r}}_i}\left( t \right)} \right)d{\mathbf{r}}} }  \nonumber \\ 
   &=  - \isum {{m_i}{{\mathbf{c}}_i}{{\mathbf{c}}_i} \cdot \frac{\partial }{{\partial {{\mathbf{r}}_i}}}{\vartheta _i}},
\end{align}
where the delta function identity $\frac{\partial }{{\partial x}}\delta \left( {x - y} \right) =  - \frac{\partial }{{\partial y}}\delta \left( {x - y} \right)$ has been used.

The derivative of the control volume function $\vartheta_i$ defined in \eq{thetadef} and \eq{rectdef} is obtained by well known manipulations of the Heaviside functions,
\begin{align}
  \frac{\partial }{{\partial {{\mathbf{r}}_i}}}{\vartheta _i} = \frac{\partial }{{\partial {{\mathbf{r}}_i}}}\vartheta \left( {\frac{{{{\mathbf{r}}_i} - {{\mathbf{r}}_m}}}{{\Delta {\mathbf{r}}}}} \right) =\;\;\;\;\;\;\;\;\;\;\;\;\;\;\;\;\;\;\;\;\;\;\;\;\;\;\;\;\;\;\;\;\;\;\;\;\;\;\;\;\;\;\;
  \nonumber \\ 
  \left[ {\delta \left( {{x_i} - {x_ - }} \right) - \delta \left( {{x_i} - {x_ + }} \right)} \right]\Pi \left( {\frac{{{y_i} - {y_m}}}{{\Delta y}}} \right)\Pi \left( {\frac{{{z_i} - {z_m}}}{{\Delta z}}} \right){\mathbf{i}} \;\; \nonumber \\ 
   + \left[ {\delta \left( {{y_i} - {y_ - }} \right) - \delta \left( {{y_i} - {y_ + }} \right)} \right]\Pi \left( {\frac{{{x_i} - {x_m}}}{{\Delta x}}} \right)\Pi \left( {\frac{{{z_i} - {z_m}}}{{\Delta z}}} \right){\mathbf{j}} \;\; \nonumber \\ 
   + \left[ {\delta \left( {{z_i} - {z_ - }} \right) - \delta \left( {{z_i} - {z_ + }} \right)} \right]\Pi \left( {\frac{{{x_i} - {x_m}}}{{\Delta x}}} \right)\Pi \left( {\frac{{{y_i} - {y_m}}}{{\Delta y}}} \right){\mathbf{k}}, \;
\end{align}
where ${\mathbf{i}}$, ${\mathbf{j}}$ and ${\mathbf{k}}$ are unit vectors in the $x$, $y$ and $z$ directions.
The physical interpretation of this function is that it selects atoms that cross the surface.
The $\mathbf{i}$ component is the $x$ surfaces of the control volume, a square in space located at the top $x_+$ or bottom $x_-$ surface bounded by $[y_-, y_+]$ and $[z_-,z_+]$. 
A similar interpretation can be applied for the $y$ and $z$ direction with the $\textbf{j}$ and $\textbf{k}$ components. 
Using the divergence theorem, we therefore have
\begin{equation}
\oint_A {{{\left( {{\mathbf{P}^K}} \right)}^T} \cdot d{\mathbf{A}}}  =  - \isum {{m_i}{{\mathbf{c}}_i}{{\mathbf{c}}_i} \cdot \frac{\partial }{{\partial {{\mathbf{r}}_i}}}{\vartheta _i}}.
\end{equation}
The left hand side, a nine-component tensor dotted with each of the six surface vectors, gives us three Cartesian components for each of the six faces of the rectangular box, making 18 terms in total. 
Each term of this expression can be matched to a corresponding term on the right hand side. 
For example, the term on the top $x$ surface of the rectangular box, $P_{xx} \Delta {A_{x + }}$ is matched to the $x_+$ face of the $\textbf{i}$ component, i.e.
\begin{align}
P_{A,xx}^K \Delta {A_{x + }}= 
\nonumber \\ 
\isum {{m_i}{c_{ix}}{c_{ix}}\delta \left( {{x_i} - {x_ + }} \right)\Pi \left( {\frac{{{y_i} - {y_m}}}{{\Delta y}}} \right)\Pi \left( {\frac{{{z_i} - {z_m}}}{{\Delta z}}} \right)}. 
\end{align}
In general, for the $\alpha \beta$ component of the surface averaged pressure tensor, with $\alpha_\pm$ denoting either top $\alpha_+$ or bottom $\alpha_-$ faces of the rectangular box, we have
\begin{align}
P_{A,\alpha \beta }^K \Delta {A_{\alpha  \pm }}= 
\nonumber \\ 
\isum {{m_i}{c_{i\beta }}{c_{i\alpha }}\delta \left( {{\alpha _i} - {\alpha _ \pm }} \right)\Pi \left( {\frac{{{\beta _i} - {\beta _m}}}{{\Delta \beta }}} \right)\Pi \left( {\frac{{{\gamma _i} - {\gamma _m}}}{{\Delta \gamma }}} \right)} .
\label{savpk}
\end{align}
To evaluate the time averages of these components of the pressure tensor, we only need to apply the delta function identity,
\begin{equation}
{\delta \left( {{\alpha _i}\left( t \right) - {\alpha _ + }} \right)} =  \sum\limits_{n} {\frac{1}{{\left| {{{\dot \alpha }_i}\left( t \right)} \right|}}\delta \left( {t - {t_n}} \right)},
\label{tsavpk}
\end{equation}
which expresses the Dirac delta function as the sum of a Dirac delta for each of the function's roots, where the index $n$ denotes the time $t_n$ at which crossings occur.
In the time averaging process, we integrate this function over time as the system evolves, and this will isolate and count molecular crossings of the $\alpha_+$ surface,
\begin{align}
\int \sum\limits_{n} {\frac{\delta \left( {t - {t_n}} \right)}{{\left| {{{\dot \alpha }_i}\left( t \right)} \right|}}} \Pi \left( {\frac{{{\beta _i(t)} - {\beta _m}}}{{\Delta \beta }}} \right)\Pi \left( {\frac{{{\gamma _i(t)} - {\gamma _m}}}{{\Delta \gamma }}} \right) dt \nonumber \\
=\sum\limits_{n} {\frac{1}{{\left| {{{\dot \alpha }_i}\left( t_n \right)} \right|}}} \Pi \left( {\frac{{{\beta _i(t_n)} - {\beta _m}}}{{\Delta \beta }}} \right)\Pi \left( {\frac{{{\gamma _i(t_n)} - {\gamma _m}}}{{\Delta \gamma }}} \right).
\label{tsavpk_int}
\end{align}
with the sifting property moving the crossing time, $t_n$, into the rectangle functions which checks if the crossing is within the area $\Delta \beta \Delta \gamma$. 
Similar expressions to \eq{tsavpk} and \eq{tsavpk_int} can be applied for the $\alpha_{-}$ surface.

To evaluate the expression for the surface average of the configurational part of the local pressure tensor, we substitute it into the volume integral of the divergence theorem, to obtain
\begin{align}
\int_{\Delta V} {\nabla  \cdot {\mathbf{P}^\phi }d{\mathbf{r}}}  = 
\nonumber \\ - \frac{1}{2}\ijsum  {\int_{\Delta V} {\nabla  \cdot \left( {{{\mathbf{r}}_{ij}}{{\mathbf{F}}_{ij}}\int_0^1 {\delta \left( {{\mathbf{r}} - {{\mathbf{r}}_s }} \right)ds } } \right)\dV} },
\end{align}
which simplifies to
\begin{align}
  \int_{\Delta V} {\nabla  \cdot {\mathbf{P}^\phi }d{\mathbf{r}}}  = 
  \nonumber \\  
  - \frac{1}{2}\ijsum  {\int_{-\infty}^{\infty} {\vartheta \left( {\frac{{{\mathbf{r}} - {{\mathbf{r}}_m}}}{{\Delta {\mathbf{r}}}}} \right)\nabla  \cdot \left( {{{\mathbf{r}}_{ij}}{{\mathbf{F}}_{ij}}\int_0^1 {\delta \left( {{\mathbf{r}} - {{\mathbf{r}}_s }} \right)ds } } \right)d{\mathbf{r}}} }   \nonumber \\ 
   = \frac{1}{2}\ijsum  {{{\mathbf{F}}_{ij}}{{\mathbf{r}}_{ij}} \cdot \int_0^1 {\frac{\partial }{{\partial {{\mathbf{r}}_s }}}{\vartheta _s }ds } }  .
\end{align}
This must be equal to the surface integral of the divergence theorem and can be written as
\begin{align}
  \oint_A {{{\left( {{\mathbf{P}^\phi }} \right)}^T} \cdot d{\mathbf{A}}} 
  =  - \frac{1}{2}\ijsum {{{\mathbf{F}}_{ij}}
  \left[ {{S_{ij}}\left( {{x_ + }} \right) - {S_{ij}}\left( {{x_ - }} \right)} \right.}    \;
  \nonumber \\[-10pt]   
   + {S_{ij}}\left( {{y_ + }} \right) - {S_{ij}}\left( {{y_ - }} \right) \;
   \nonumber \\
  \left. { + {S_{ij}}\left( {{z_ + }} \right) - {S_{ij}}\left( {{z_ - }} \right)} \right],
\end{align}
where
\begin{align}
{S_{ij}}\left( {{\alpha _ + }} \right) \equiv \frac{1}{2}\left[ {\operatorname{sgn} \left( {{\alpha _ + } - {\alpha _i}} \right) - \operatorname{sgn} \left( {{\alpha _ + } - {\alpha _j}} \right)} \right]
\nonumber \\
\times \Pi \left( {\frac{{{\beta _{\alpha  + }} - {\beta _m}}}{{\Delta \beta }}} \right)\Pi \left( {\frac{{{\gamma _{\alpha  + }} - {\gamma _m}}}{{\Delta \gamma }}} \right).
\label{Sij}
\end{align}
The final equation for practical computation of the $\alpha \beta$ component of the pressure tensor at the $\alpha_+$ surface of the rectangular box is
\begin{equation}
\begin{split}
P_{A, \alpha \beta }^\phi \Delta {A_{\alpha  + }} = \;\;\;\;\;\;\;\;\;\;\;\;\;\;\;\;\;\;\;\;\;\;\;\;\;\;\;\;\;\;\;\;\;\;\;\;\;\;\;\;\;\;\;\;\;\;\;\;\;\;\;\;\;\;\;\;\;\;\;
\nonumber \\  
- \frac{1}{4}\ijsum {{F_{\beta ij}}\left[ {\operatorname{sgn} \left( {{\alpha _ + } - {\alpha _i}} \right) - \operatorname{sgn} \left( {{\alpha _ + } - {\alpha _j}} \right)} \right] } \\
\times {{\Pi \left( {\frac{{{\beta _{\alpha  + }} - {\beta _m}}}{{\Delta \beta }}} \right)\Pi \left( {\frac{{{\gamma _{\alpha  + }} - {\gamma _m}}}{{\Delta \gamma }}} \right)} }, 
\end{split}
\end{equation}
where $\beta_{\alpha +} \equiv \beta_+ - \beta_i - \frac{\beta_{ij}}{ \alpha_{ij}} (\alpha_+ - \alpha_i)$.
The two rectangle functions select only those $ij$ interactions whose $\mathbf{r}_{ij}$ vector intersects the $\alpha _ +$ surface within the area $\Delta \beta \Delta \gamma$ and the sgn function selects only those interactions having one particle on each side of the $\alpha _ +$ surface. In the limit of infinite planar surfaces, the last two rectangle functions are always equal to 1, since $\Delta \beta $ and $\Delta \gamma$ both become infinitely large, and the expression reduces to the method of planes result. A similar treatment can be applied to the $\alpha_{-}$ surface.

The corresponding version of the divergence theorem for the heat flux vector is
\begin{equation}
\oint_A {{{\mathbf{J}}_q} \cdot d{\mathbf{A}}}  = \int_{\Delta V} {\nabla  \cdot {{\mathbf{J}}_q}d{\mathbf{r}}} .
\end{equation}
The expression for the surface averaged kinetic part of the heat flux vector is analogous to the corresponding part of the pressure tensor,
\begin{align}
J_{qA,\alpha }^K \Delta {A_\alpha }= \;\;\;\;\;\;\;\;\;\;\;\;\;\;\;\;\;\;\;\;\;\;\;\;\;\;\;\;\;\;\;\;\;\;\;\;\;\;\;\;\;\;\;\;\;\;\;\;\;\;\;\;\;\;\;\;\;\;\;
\nonumber \\
\isum {{u_i}{c_{i\alpha }}\delta \left( {{\alpha _i} - {\alpha _ + }} \right)\Pi \left( {\frac{{{\beta _i} - {\beta _m}}}{{\Delta \beta }}} \right)\Pi \left( {\frac{{{\gamma _i} - {\gamma _m}}}{{\Delta \gamma }}} \right)} .
\end{align}
The configurational part of the heat flux vector is more complicated. Separating the particle velocity and fluid streaming velocity terms of the configurational part of \eq{jqlocal} we have
\begin{align}
  \int_{\Delta V} {\nabla  \cdot {{\mathbf{J}}^\phi }d{\mathbf{r}}}  =   \;\;\;\;\;\;\;\;\;\;\;\;\;\;\;\;\;\;\;\;\;\;\;\;\;\;\;\;\;\;\;\;\;\;\;\;\;\;\;\;\;\;\;\;\;\;\;\;\;\;\;\;\;\;\;\;
\nonumber \\
- \frac{1}{2}\ijsum {{{\mathbf{r}}_{ij}}{{\mathbf{F}}_{ij}}:{{\mathbf{v}}_i}\int_0^1 {\int_{\Delta V} {\nabla \delta \left( {{\mathbf{r}} - {{\mathbf{r}}_s }} \right)d{\mathbf{r}}} ds } }  \nonumber \\ 
   + \frac{1}{2}\ijsum {{{\mathbf{F}}_{ij}}{{\mathbf{r}}_{ij}}:\int_0^1 {\int_{\Delta V} {\nabla \left[ {{\mathbf{v}}\left( {\mathbf{r}} \right)\delta \left( {{\mathbf{r}} - {{\mathbf{r}}_s }} \right)} \right]d{\mathbf{r}}} ds } }.  
   \label{Integrated_energy}
\end{align}
The first term is analogous to the configurational part of the pressure tensor and results in
\begin{align}
   - \frac{1}{2}\ijsum {{{\mathbf{r}}_{ij}}{{\mathbf{F}}_{ij}}:{{\mathbf{v}}_i}\int_0^1 {\int_{\Delta V} {\nabla \delta \left( {{\mathbf{r}} - {{\mathbf{r}}_s }} \right)d{\mathbf{r}}} ds } } 
   \nonumber \\
   =  - \frac{1}{2}\ijsum {{{\mathbf{F}}_{ij}} \cdot {{\mathbf{v}}_i}
   \left[  {{S_{ij}}\left( {{x_ + }} \right) - {S_{ij}}\left( {{x_ - }} \right)} \right.}  \;
   \nonumber \\[-10pt]
   + {{S_{ij}}\left( {{y_ + }} \right) - {S_{ij}}\left( {{y_ - }} \right)} \;
   \nonumber \\ 
  \left. { +  {{S_{ij}}\left( {{z_ + }} \right) - {S_{ij}}\left( {{z_ - }} \right)} } \right]
  \label{1st_config_term}
\end{align}
with $S_{ij}$ again defined by \eq{Sij}.
The second term in \eq{Integrated_energy}, containing the fluid streaming velocity, is
\begin{equation}
\begin{split}
  \frac{1}{2}\ijsum {{{\mathbf{F}}_{ij}}{{\mathbf{r}}_{ij}}:\int_0^1 {\int_{\Delta V} {\nabla \left[ {{\mathbf{v}}\left( {\mathbf{r}} \right)\delta \left( {{\mathbf{r}} - {{\mathbf{r}}_s }} \right)} \right]d{\mathbf{r}}} ds } }   \\ 
   =  - \frac{1}{2}\ijsum {{{\mathbf{F}}_{ij}}{{\mathbf{r}}_{ij}}: \! \int_{ - \infty }^\infty \!\!\!\! {\Pi \left( {s  - \frac{1}{2}} \right)\frac{\partial }{{\partial {{\mathbf{r}}_s }}}\vartheta \left( {\frac{{{{\mathbf{r}}_s } - {{\mathbf{r}}_m}}}{{\Delta {\mathbf{r}}}}} \right){\mathbf{v}}\left( {{{\mathbf{r}}_s }} \right)ds } } 
\end{split}
\end{equation}
which can be integrated to give
 \begin{align}
- \frac{1}{2}\ijsum {{{\mathbf{F}}_{ij}} \cdot \left[ {\left( {{S_{ij}}\left( {{x_ - }} \right){\mathbf{v}}\left( {{{\mathbf{r}}_{x - }}} \right) - {S_{ij}}\left( {{x_ + }} \right){\mathbf{v}}\left( {{{\mathbf{r}}_{x + }}} \right)} \right)} \right.}  \;\;\,
\nonumber \\[-10pt]  
   + \left( {{S_{ij}}\left( {{y_ - }} \right){\mathbf{v}}\left( {{{\mathbf{r}}_{y - }}} \right) - {S_{ij}}\left( {{y_ + }} \right){\mathbf{v}}\left( {{{\mathbf{r}}_{y + }}} \right)} \right) \;\;\,
\nonumber \\
  \left. { + \left( {{S_{ij}}\left( {{z_ - }} \right){\mathbf{v}}\left( {{{\mathbf{r}}_{z - }}} \right) - {S_{ij}}\left( {{z_ + }} \right){\mathbf{v}}\left( {{{\mathbf{r}}_{z + }}} \right)} \right)} \right],  \label{2nd_config_term}
\end{align}
where ${\mathbf{v}}\left( {{{\mathbf{r}}_{x + }}} \right) = {\mathbf{v}}\left( {{{\mathbf{r}}_i} + \frac{{{x_ + } - {x_i}}}{{{x_{ij}}}}{{\mathbf{r}}_i}} \right)$ is the fluid velocity at the point where the $\mathbf{r}_{ij}$ vector intersects the $x_ +$ plane and similarly for ${\mathbf{v}}\left( {{{\mathbf{r}}_{x - }}} \right)$, etc.

Inserting \eq{1st_config_term} and \eq{2nd_config_term} into \eq{Integrated_energy} and using the divergence theorem for the left hand side, we find that the final result for practical computation of the surface heat flux vector is
\begin{align}
  J_{qA,\alpha }^\phi \Delta {A_{\alpha  + }} 
   =  - \frac{1}{2}\ijsum {{F_{\beta ij}}{v_{\beta i}}{S_{ij}}\left( {{\alpha _ + }} \right)}
      \nonumber \\ 
+ \frac{1}{2}\ijsum {{F_{\beta ij}}{v_\beta }\left( {{{\mathbf{r}}_{\alpha + }}} \right){S_{ij}}\left( {{\alpha _ + }} \right)}  
   \nonumber \\ 
   =  - \frac{1}{2}\ijsum {{F_{\beta ij}}\left( {{v_{\beta i}} - {v_\beta }\left( {{{\mathbf{r}}_{\alpha  + }}} \right)} \right){S_{ij}}\left( {{\alpha _ + }} \right)}. 
   \label{CV_Heat_flux}
\end{align}
This result shows that the relative velocity that appears in the expression for the surface averaged configurational part of the heat flux vector is the particle velocity minus the streaming velocity at the point where the $\mathbf{r}_{ij}$ vector intersects the $\alpha_ +$ plane. This is again the plane peculiar velocity discussed by \citet{Todd_Evans_Daivis_95} Again, a similar expression exists for the $\alpha_{-}$ plane.

As with the volume average form, to calculate the surface averaged heat flux in a simulation, we decompose $\oint_A \heatflux \cdot d\textbf{A} = \oint_A \left( \heatflux^{\!\!\! K} + \heatflux^{\!\!\!\! \phi} \right) \cdot d\textbf{A} $ into kinetic and configurational parts, obtaining the kinetic component from
\begin{align}
J_{qA,x}^K = \frac{1}{{\Delta {A_x}}} \displaystyle\sum_{i=1}^N {e_i}\left( {{v_{ix}} - {{\bar v}_x}\left( {{r_m}} \right)} \right)\delta \left( {{x_i} - {x_ + }} \right)   
\nonumber \\
\times \Pi \left( {\frac{{{y_i} - {y_m}}}{{\Delta y}}} \right)\Pi \left( {\frac{{{z_i} - {z_m}}}{{\Delta z}}} \right)
\label{MOP_advection}
\end{align}
and the configurational component from
\begin{align}
J_{qA,x}^\phi  =  - \frac{1}{{\Delta {A_{x + }}}}\frac{1}{2}\ijsum {{{\mathbf{F}}_{ij}} \cdot \left( {{{\mathbf{v}}_i} - \bar{\mathbf{v}}\left( {{{\mathbf{r}}_{x + }}} \right)} \right){S_{ij}}\left( {{x_ + }} \right)}, 
\label{MOP_stressheat}
\end{align}
where velocity $\bar{\mathbf{v}}$ is again from definition \eq{vel_definition} extrapolated to the surface of the cell. 
This can again be computed in two parts that are combined at the end of the run.
We have shown only the $x$ component, with similar forms for the $y$ and $z$ component heat fluxes on the other surfaces of the volume. 

\section{Simulations}
\label{sec:Simulation}

In this work, we use two different types of molecular simulation in order to isolate the three coefficients of thermal conductivity given in \eq{tc_tensor}. Both cases simulate a wall bounded channel. In system 1, we apply a temperature gradient to obtain the Fourier's law thermal conductivity coefficient $\lambda$ in the absence of a shear rate, recording heat flux over a range of densities. These heat flux results are compared to experimental results over a similar range of densities. In system 2, we simulate wall driven Couette flow over the same range of densities where we observed good experimental agreement for system 1. We obtain the coefficients of strain rate dependent thermal conductivity both parallel to the flow, $\lambda_1$, and normal to the flow, $\lambda_2$. In this section, details of the molecular simulation methodology common to both systems are first outlined before the key features of systems 1 and 2 are discussed. 

\subsection{Setup}
\label{sec:setup}

In this work, we use a simple truncated and shifted Lennard-Jones potential (the so-called Weeks-Chandler-Anderson (WCA) potential \cite{Weeks71}) in order to test the theoretical predictions: 
\begin{align}
  \phi _{ij}^{WCA} = \left\{ \begin{array}{l}
4\epsilon \left[ {{{\left( {\frac{\sigma }{{{r_{ij}}}}} \right)}^{12}} - {{\left( {\frac{\sigma }{{{r_{ij}}}}} \right)}^6}} \right] + \epsilon \quad {r_{ij}} < {2^{{1 \mathord{\left/
 {\vphantom {1 6}} \right.
 \kern-\nulldelimiterspace} 6}}}\sigma \\
0\quad \quad \quad \quad \quad \quad \quad \quad \quad \quad \quad \;\,{r_{ij}} \ge {2^{{1 \mathord{\left/
 {\vphantom {1 6}} \right.
 \kern-\nulldelimiterspace} 6}}}\sigma .
\end{array} \right.
  \label{LJ_potential}
\end{align}
More realistic potentials and fluids have been avoided here because they could introduce further complexity due to rotational, non-Newtonian and coupled transport effects which might obscure the phenomena that are the focus of this work.
The Lennard-Jones potential also has the advantage that it gives good agreement with experiments using liquid Argon. 
The equations of motion for the fluid are solved using the Leapfrog-Verlet scheme.
The WCA cut off distance, $r_c = 2^{\frac{1}{6}}$, was used, an assumption justified by previous observations that many transport properties are dominated by repulsive interactions \citep{todd_daivis_2017}.
Both systems have a timestep of $\Delta t = 0.005$. Each simulation ran for a total of $1,000,000$ steps with samples taken every $25$ timesteps and results were written to disk every 25 samples (\ie an output every $625$ timesteps).
The system was initialised to an integer number of FCC units with the liquid region created by randomly removing atoms until the desired liquid density, $\rho_l$, was obtained.
The walls remained as an FCC crystal lattice with density of $\rho_w = 1.0$ and a thickness of $4$ atomic units ($L_{wall}=4$) from the top and bottom boundaries in the $y$ direction. 
The walls were tethered using the anharmonic potential from \citet{Petravic_Harrowell} with spring constants for second, fourth and sixth power of displacement terms $k_2=0.0, k_4=5\times10^3$ and $k_6=5\times10^6$ respectively.
This potential was chosen by \citet{Petravic_Harrowell} to allow sufficient motion for momentum and energy interchange while preventing wall atoms from moving far enough to compromise heat flux or shear stress definitions. 
The system temperature was set to $T=1.2$ initially with the temperature of both walls controlled using a Nos\'{e}-Hoover thermostat applied only to the outer $L_{therm} = 2$ reduced units of atoms in $y$, \ie{} half the wall is thermostatted in $y$ with the $2$ units closest to the liquid unthermostatted.
The anharmonic tethering is purported to have better thermostatting characteristics, with a harmonic potential shown to impact the canonical distribution achieved by the Nos\'{e}-Hoover thermostat \citep{Petravic_Harrowell_2005}.

The wall atoms have the following equations of motion:
\begin{subequations}
\begin{eqnarray}
\MDvel  &=& \MDpvel + v_w {\bf{\hat{n}}}_x, \\
m_i \MDaccel  &=& \textbf{F}_{i} + \textbf{F}_{i}^{teth} - \psi m_i \MDpvel, \\
\textbf{F}_{i}^{teth} &=& \textbf{r}_{i_0}  \left( 4 k_{4} r_{i_0}^{2}+6 k_{6} r_{i_0}^{4} \right), \\
\dot{\psi} &=& \frac{1}{Q_\psi} \left[  \displaystyle\sum_{i=1}^{N_{_{therm}}} m_i \MDpvel \cdot \MDpvel - 3T_0 \right]. 
\label{NH_verify} 
\end{eqnarray}
\end{subequations}
Here $Q_\psi$ is a coefficient which determines the Nos\'{e}-Hoover thermostat's heat bath inertia, namely how strongly the wall temperature will tend to set point $T_0$. $\textbf{r}_{i_{0}}=\textbf{r}_{i}-\textbf{r}_{0}$, is atom $i$'s position relative to lattice site $\textbf{r}_{0}$ (the atom's tethering location) where it slides at the wall speed $v_w$ in the streamwise $x$ direction with unit vector ${\bf{\hat{n}}}_x = [1, 0, 0]$. The two walls are thermostatted independently, \ie{} they are attached to separate heat baths where the strength of thermostatting is chosen to be proportional to $N_{therm}$, the number of atoms thermostatted in each wall, with $Q_\psi \define 5\times10^{-4} N_{therm}$.

 \begin{figure}
\includegraphics[width=0.48\textwidth]{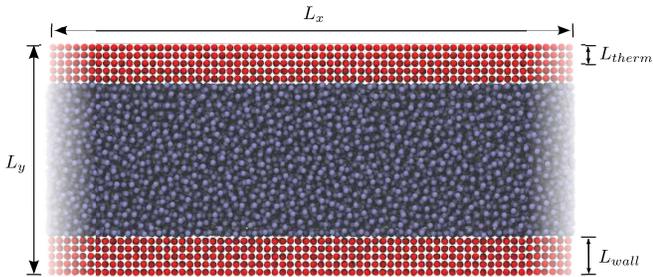}
\caption{Channel schematic with tethered wall atoms in red and liquid atoms in blue, with extents of system denoted by $L_x$, $L_y$, wall size by $L_{wall}$ and thermostatted region by $L_{therm}$ which is offset from the liquid region and applied on both top and bottom walls.}  
\label{schematic}
\end{figure}

Before averages are taken, the simulation is run for an initial period of time to reach steady state, which for system 1 is indicated by a time-independent approximately linear temperature profile, while in system 2 it is indicated by time-independent, approximately linear velocity and parabolic temperature profiles.

\subsubsection{System 1}
\label{sec:system1}

In system 1, we used separately thermostatted walls and densities varying from $\rho_l=0.4$ to $0.9$ with  $T_{bot} = 0.8$ at the bottom and $T_{top} = 1.2$ at the top, giving a liquid temperature of $T_l=1.0$ on average.
A range of average temperatures were also studied at fixed densities $\rho_l= \{ 0.65, 0.75, 0.85 \}$, varying from $T_{bot} = 0.5$ and $T_{top} = 0.9$ to $T_{bot} = 1.0$ and $T_{top} = 1.4$ so that the temperature gradient remained constant.
The system size is $L_x = 31.75$, $L_y =  47.62$ and $L_z =   31.75 $ which results in a liquid height of $39.62$, which for the case with $\rho_{l}=0.8$ has $N = 37,902$ atoms, but varying for other densities. The domain is split into $22$ by $66$ by $22$ averaging bins in $x,y$ and $z$ respectively. 

\subsubsection{System 2}
\label{sec:system2}

In system 2, as depicted in Fig. \ref{schematic}, both walls are thermostatted to the same temperature $T_{wall} = 0.7$ with the walls atoms counter sliding at velocities set to $v_w=\pm 1$. 
The wall temperature was chosen to give a liquid temperature of $T_l=1$ on average.
In order to give higher shear rates, a smaller channel height, $L_y$, was used than in the temperature study of system 1. 
As a result, the shear dependent terms in equations \eq{approx_qx} and \eq{approx_qy} are enhanced but with height chosen to give a liquid region of $15.81$, which is the smallest size where bulk behaviour has been shown to broadly match hydrodynamics \citep{Travis_Gubbins_00}. 
The domain size is expanded in other directions to improve statistics, with $L_x = 139.69$, $L_y =  23.81$ and $L_z = 93.66$ so the number of atoms modelled is $N = 259,359$ for the case with a density of $\rho_{l}=0.8$, again varying for other densities. 
The domain is split into $12$ by $256$ by $8$ averaging bins in $x,y$ and $z$ respectively.

\section{Results and Discussion}
\label{sec:Results}

In this Section, the results from the molecular simulations are presented and discussed, starting with system 1 in section \ref{sec:results_sys1} and then system 2 in section \ref{sec:results_sys2}. 
Results from system 1 are presented first in order to obtain the thermal conductivity coefficient in the absence of shear.

\begin{figure}
\includegraphics[width=0.48\textwidth]{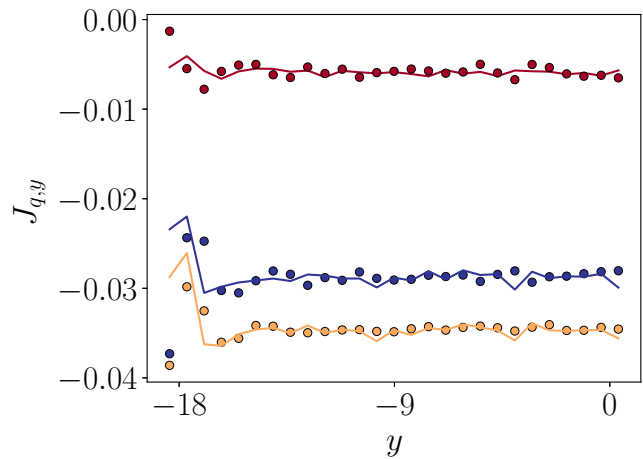}
\caption{Terms which contribute to the heat flux at $\rho_l=0.6$, with MOP kinetic term $J_{qA,y}^K $ as {red} circles, configurational term $J_{qA,y}^\phi$ shown by blue circles and the sum of both $J_{qA,y}$ displayed as {gold} circles. The VA kinetic term is shown as a {red} line, $J_{qV,y}^K$, configurational term as a blue line $J_{qV,y}^\phi$ and the sum of both is shown as a {gold} line $J_{qV,y}$.} 
\label{diff_fourier_energy_terms}
\end{figure}

\subsection{System 1 -- Temperature Gradient}
\label{sec:results_sys1}

As outlined in Section \ref{sec:system1}, a temperature gradient was induced by thermostatting each of the atomistic walls separately to different temperatures. The computed heat fluxes as a function of channel position are presented in Fig. \ref{diff_fourier_energy_terms}, where we have shown the configurational component, $J_{q,y}^{\phi}$, kinetic component, $J_{q,y}^{K}$, and the resulting total heat flux calculated from the sum of both, $J_{q,y}=J_{q,y}^{K}+J_{q,y}^{\phi}$. The resulting heat flux is calculated for the channel using using \eq{VA_advection}/\eq{MOP_advection} for VA/MOP kinetic contributions and \eq{VA_stressheat}/\eq{MOP_stressheat} for VA/MOP configurational components. Fig. \ref{diff_fourier_energy_terms} shows results from both the VA and MOP approaches to calculating the heat flux, split into kinetic and configurational parts, with good agreement between all computations. The near-wall discrepancy between VA and MOP is attributed to density oscillations which are measured differently 
by the two methods, either as atoms inside a volume or crossings over a plane.

We obtain the temperature gradient $\partial T / \partial y$ from the derivative of a linear fit to temperature in the inner liquid region of the channel, starting $2.5$ reduced units away from the wall to avoid density oscillations and the Kapitza resistance at the surface. 
The heat flux, $J_{q,y}$, measured for the same inner liquid region, is then divided by temperature gradient to obtain the liquid's thermal conductivity directly, $ \lambda = - J_{q,y}/(\partial T / \partial y)$. 
Only the thermal conductivity of Fourier's law $\lambda$, is relevant as the strain rate of system 1 is zero, so \eq{Jq_constit} and \eq{tc_tensor} simplify to
\begin{eqnarray}
J_{q,y}^{approx} &=& - \lambda \frac{\partial T}{\partial y}.
\label{Fourier_Heat_flux_only}
\end{eqnarray}
This process is applied over the range of different liquid densities from $\rho_l = 0.4$ to $\rho_l = 0.95$.
\begin{figure}
\includegraphics[width=0.48\textwidth]{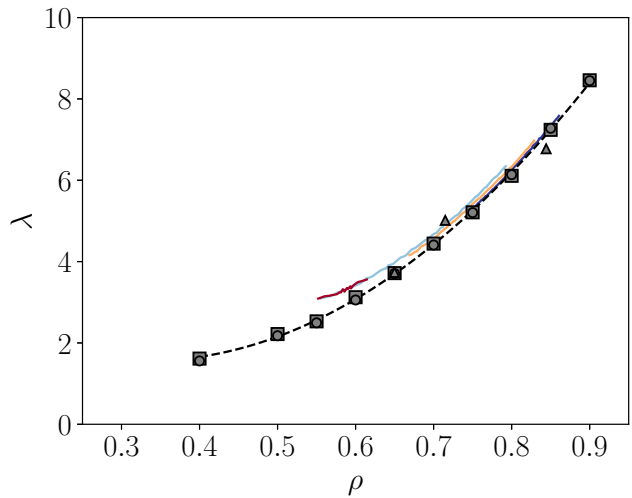}
\caption{Summary of MD density study. VA square and MOP circles with lines showing experimental results, blue $110\text{K}$, gold $124\text{K}$, red $138\text{K}$ and light blue $140\text{K}$.}  
\label{Fourier_study_k}
\end{figure}
The resulting thermal conductivity as a function of density is presented in Fig. \ref{Fourier_study_k} and compared to experimental results for liquid Argon from \citet{Roder1987} with both axes expressed in reduced units.

Good agreement is observed over the experimental range with discrepancies only apparent for $\rho_l$ below $0.6$ and above $0.85$. This is attributed to the limitations of the WCA model to reproduce experimental results for liquid Argon at more extreme densities. As a result, we limited the study to densities, $\rho_l$, in the range $[0.6, 0.85]$ in the study of planar shear flow in system 2.

\begin{figure}
\includegraphics[width=0.48\textwidth]{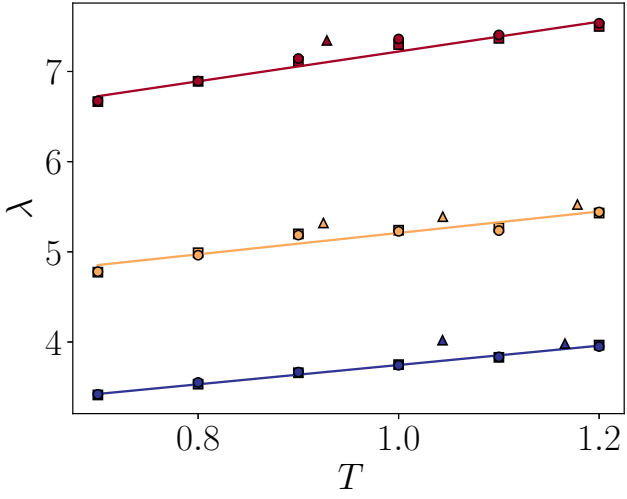}
\caption{Summary of MD temperature study. VA squares, MOP circles and triangles showing the relevant experimental results, with three densities in reduced units, blue $\rho_l=0.65$, gold $\rho_l=0.75$, red $\rho_l=0.85$ with numerical results fitted using straight lines with gradients $m_T$ of $1.07$, $1.19$ and $1.57$ respectively.}  
\label{Fourier_study_temperature_k}
\end{figure}

The study over a range of density was performed at a constant mean temperature of $T_l=1.0$, however the Fourier's law thermal conductivity is also weakly dependent on temperature.
In system 2, the mean liquid temperature was only controlled indirectly through the wall thermostats and can vary as shear rate changes. To quantify this, we performed a study between $T=0.7$ and $T=1.2$ at three of the densities which agree with experimental results $\rho = \{0.65, 0.75, 0.85\}$, shown in Fig \ref{Fourier_study_temperature_k}.
The experimental data from \citet{Roder1987} at comparable temperature and density is limited but all relevant values are included on Fig \ref{Fourier_study_temperature_k} and agree fairly well.
The simulation results in Fig \ref{Fourier_study_temperature_k} are fitted using a linear function for each density. This temperature correction to the Fourier's law thermal conductivity can then be included into a general fit, $\lambda_f$,
\begin{align}
\lambda_f(\rho_l, T_l) = \lambda_f^\prime(\rho_l) + m_T(\rho_l) \left(T_l-1\right)
\label{sys1_lambda}
\end{align}
where $T_l$ is the mean measured liquid temperature and $m_T=m_T(\rho_l)$ is a gradient fitted to data at each of the three density values shown in Fig \ref{Fourier_study_temperature_k}. 
Using second order polynomial fits to the data in Fig. \ref{Fourier_study_k} gives $\lambda_f^\prime(\rho_l) = 21.3 \rho_l^2 -14.2 \rho_l + 3.92 $ and fitting a second order curve to the varying gradients obtained from Fig \ref{Fourier_study_temperature_k} yields $m_T(\rho_l) = 13.0 \rho_l^2 -17.0 \rho_l + 6.63$ which allows $\lambda_f(\rho_l, T_l)$ to be approximated for a given temperature and density.

\subsection{System 2 -- Velocity Gradient}
\label{sec:results_sys2}

\begin{figure}
\includegraphics[width=0.48\textwidth]{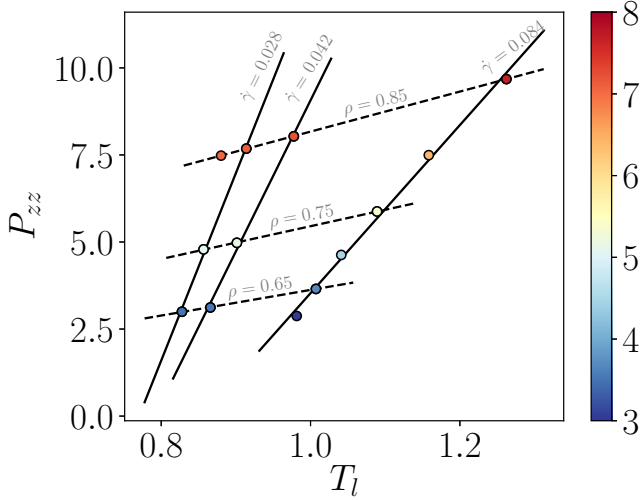}
\caption{Overview of the average liquid temperature, $T_l$ against spanwise pressure $P_{zz}$ for all cases run in this section, with dotted lines showing studies with varying strain rate and constant density (isochores) while solid lines show studies with constant applied strain rate and varying density. Points are coloured using the fit given in \eq{sys1_lambda} to estimate the Fourier's law coefficient $\lambda$ for each temperature and density values.}
\label{TvsP}
\end{figure}

Having obtained the thermal conductivity in the absence of shear, we move on to system 2 as outlined in Section \ref{sec:system2}, i.e. planar shear flow. We ran simulations over a range of strain rates by changing system size, $L_y = \{23.8, 47.5, 71.3, 95.2\}$, as well as varying over the range of densities that matched to experimental results in system 1, $\rho_l = \{0.6,0.65,0.7,0.75,0.8,0.85\}$.
The challenge in varying density and channel heights is that the system temperature and pressure also change in an uncontrolled manner, outlined in Fig \ref{TvsP}, which summarises all the cases run in this section.
The lines of constant density and constant applied strain rate are shown on Fig \ref{TvsP}, highlighting that despite only varying either density or strain rate, there results a corresponding change in temperature and pressure.
The Fourier's law thermal conductivity coefficient, $\lambda$, at each temperature and density is estimated from \eq{sys1_lambda} and shown in Fig \ref{TvsP} by the colouring of the points.
This variation is observed despite thermostatting both walls to a temperature $T_{wall} = 0.7$ in all cases.

\begin{figure}
\includegraphics[width=0.48\textwidth]{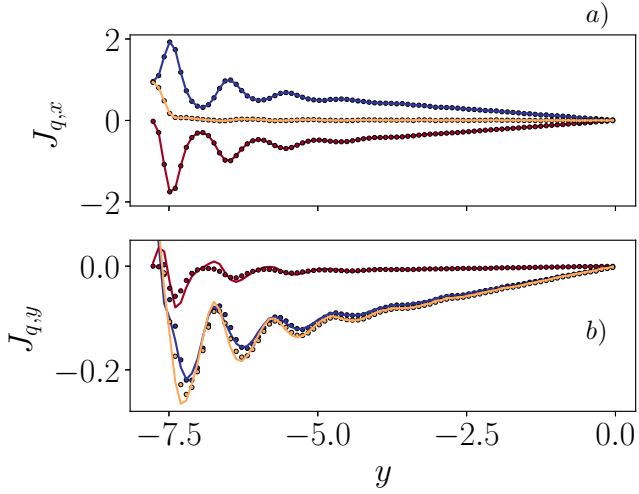}
 \put(-25,180){$a)$}
 \put(-25,60){$b)$}
\caption{Terms which contribute to the heat flux at a liquid density of $\rho_l=0.75$. The $\alpha=x$ components are shown in (a) and $\alpha=y$ components in (b) with the MOP kinetic term $J_{qA,\alpha}^K $ as red circles and configurational flux $J_{qA,\alpha}^\phi$ as blue circles with sum of both $J_{qA,\alpha}$ shown as gold circles. The VA kinetic contribution is shown as a red line, $J_{qV,\alpha}^K$, configurational contribution as a blue line, $J_{qV,\alpha}^\phi$, and the sum of both displayed as a gold line, $J_{qV,\alpha}$.}
\label{diff_shear_energy_terms}
\end{figure}

The method for obtaining the components of heat flux is identical to the one used in system 1, based on \eq{VA_advection}/\eq{MOP_advection} for kinetic contributions and \eq{VA_stressheat}/\eq{MOP_stressheat} for configurational.
However, in the presence of shear flow, we have additional components of heat flux. The $x$ component of heat flux, $J_{q,x}$, as a function of channel position is shown in Fig. \ref{diff_shear_energy_terms} (a) and $y$ component of heat flux $J_{q,y}$ is shown in Fig. \ref{diff_shear_energy_terms} (b). These are decomposed into the various contributions as before: kinetic heat flux, $\heatflux^{\!\! K}$, and configurational flux, $\heatflux^{\!\! \phi}$ as well as their sum $\heatflux{}$.  All simulations were run until the thermal and hydrodynamic steady state has been reached before properties were accumulated. 

Both the VA and MOP approaches are used to calculate the heat flux and good agreement is observed between the two measurement techniques, as shown in Fig. \ref{diff_shear_energy_terms}. Previous studies have shown that the VA and MOP stresses are equivalent \citep{Heyes_et_al} in the small bin limit and this observation is extended here to the energy equation and heat flux. The MOP form of stress is known to have worse statistics than VA \citep{Hadjiconstantinou_et_al, Smith_et_al17}, a problem which is even more acute in the calculation of heat flux. Short averaging periods ($25$ records collected over $625$ timesteps) were found to result in a systematic error in the $J_{q,x}$ component of the MOP heat flux (when compared to the VA value). This may be due to very long correlations between $\Pressuretype$ and $\CFDvel$. This error was found to disappear when averages from the entire one million timesteps simulation are used. 
In order to ensure agreement between MOP and VA approaches, averages of velocity, $\CFDvel$, and the pressure tensor, $\Pressuretype$, are then taken over the entire simulation run length to calculate the correct stress heating term, $\Pressuretype \cdot \CFDvel$. Interpolation of density and velocity measurements to the appropriate cell surface are also used to improve agreement between VA and MOP (the so-called `plane peculiar' velocity problem as noted before and first pointed out by Todd et al. \cite{Todd_Evans_Daivis_95}). These are required as density and velocity measurements are obtained for atoms in a volume, so they represent the volume average about the centre of that cell, inducing a systematic error of order $\Delta y/2$ when calculating quantities at the cell surfaces. Having obtained the profiles for $J_{q,x}$ and $J_{q,y}$ in the sheared channel of system 2, we turn our attention to the calculation of the thermal conductivity coefficients $\lambda_1$ and $\lambda_2$.

For the case of planar shear flow with constant strain rate, \eq{Jq_constit} and \eq{tc_tensor} can be simplified to,
\begin{eqnarray}
J_{q,x}^{approx}  &=& \Lambda_1 \frac{\partial T}{\partial y} \;\;\;\;\;\;\;\; \;\;\;\;\;\;\; \;\;\;\;\;\;\;\;\;
\label{approx_qx} \\
J_{q,y}^{approx} &=& - \lambda \frac{\partial T}{\partial y}  - \Lambda_2\frac{\partial T}{\partial y},
\label{approx_qy}
\end{eqnarray}
where we have collected the strain rate dependent terms into single coefficients $\Lambda_1 \define \lambda_1 \dot{\gamma}$ and $\Lambda_2 \define 3 \lambda_2 \dot{\gamma}^2$.
In the following section, we aim to use the values for $J_{q,x}$ and $J_{q,y}$ measured in the molecular simulation to obtain the two unknown thermal conductivity coefficients.

\begin{figure*}
\includegraphics[width=\textwidth]{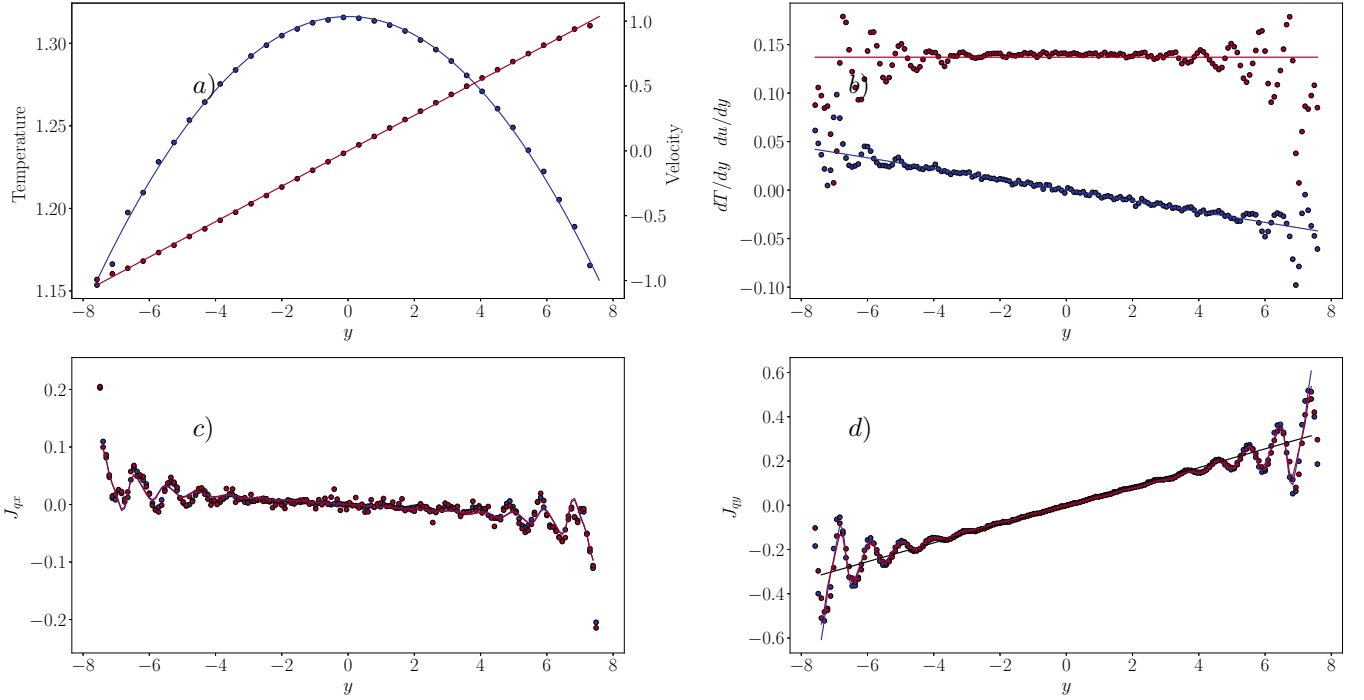}
 \put(-438,230){$a)$}
 \put(-190,230){$b)$}
 \put(-438,100){$c)$}
 \put(-190,100){$d)$}
\caption{Summary of the method used to get coefficients for $\heatflux{}^{approx}$ at $\rho_l=0.85$.  (a) Temperature $T$ is shown in blue and streamwise ($x$) velocity $v_x\left(y\right)$ shown in red. Circles are MD results (displayed every $5$ cells) and lines are least squares fits to the data. (b) Lines show derivatives of the least squares fits and circles show the numerical derivatives of the MD results. (c) VA $J_{q,x}$ displayed by blue circles and MOP values shown in red circles, with lines depicting the least square fit to $J_{q,x}^{approx}$ from \eq{approx_qx_grho}. (d) $J_{q,y}$ again with blue circles for VA, red circles for MOP, fits to $J_{q,y}^{approx}$ from \eq{approx_qy_grho} and a black line showing Fourier's law, $-\lambda dT/dy$, with coefficient obtained from the temperature gradient study.}  
\label{shear_heatflux_terms}
\end{figure*}

In order to obtain the coefficients $\Lambda_1$ and $\Lambda_2$, the expressions for $\partial T/\partial y$ and $\dot{\gamma} = \partial v / \partial y$ must be obtained.
The near-wall density oscillations in the channel of system 2 affect the velocity and temperature results and prove especially problematic in calculating the numerical derivatives required for $\dot{\gamma} = \partial v / \partial y$ and $\partial T / \partial y$. In addition to density oscillations, numerical derivatives are also susceptible to noise in the data, which persist even for the large numbers of samples collected in these studies. 
To avoid these problems, least squares fits to the parabolic temperature profile and linear velocity functions are used and the resulting functional forms are differentiated instead. 
The profiles of velocity and temperature are shown in Fig. \ref{shear_heatflux_terms} (a) with fits to linear $v^{LS} =  a_1 y$ and quadratic, $T^{LS} = b_1 y^2 + b_2$, functions respectively, where $\{a_1,b_1, b_2 \}$ are fitting parameters. 
The expressions, $dT^{LS}/dy$ and $\partial v^{LS} / \partial y = \dot{\gamma}^{LS}$, shown in Fig. \ref{shear_heatflux_terms} (b) as lines are obtained from the analytical derivative of the least squares fits.
In doing this, we have decomposed the temperature profile, $T = T^{LS} + T^\prime$, into a purely quadratic part $ T^{LS}$, as well as a separate density and noise dependent part $T^\prime$. 
Although the parabolic fit is good in Fig \ref{shear_heatflux_terms} (a), departures due to both noise and density stacking are evident in the derivative of Fig \ref{shear_heatflux_terms} (b).
The derivative of the component of temperature due to density oscillations and noise can be expressed mathematically as $ \partial T^\prime / \partial y = \partial T / \partial y  - \partial T^{LS} /\partial y$.
We replace $\partial T^\prime / \partial y$ with the less noisy gradient of density, $\partial T^\prime / \partial y \to \partial \rho/\partial y$, to improve the quality of the fit discussed in Appendix  \ref{sec:error}.
\begin{figure}
\includegraphics[width=0.48\textwidth]{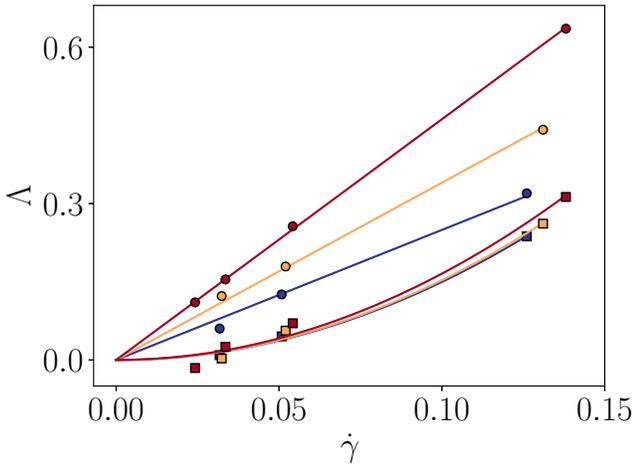}
\caption{Volume average heat flux terms plotted as a function of strain rate at densities: (blue) $\rho_l = 0.65$, (gold) $\rho_l = 0.75$ and (red) $\rho_l = 0.85$, shown by circles for parallel ($x$-component) heat flux $\Lambda_1$ with a linear best fit going through zero, and normal ($y$-component) heat flux $\Lambda_2$ is shown using squares, obtained by shifting to ensure heat flux is zero at a strain rate of zero for each density and fitted with a quadratic best fit.}  
\label{qx_qy2_vs_strain}
\end{figure}

The equations used in the fitting shown in Fig \ref{shear_heatflux_terms} $c)$ and $d)$ are therefore,
\begin{align}
J_{q,x}^{approx}  = \Lambda_1 \frac{\partial T^{LS}}{\partial y} + d_x\frac{\partial T^\prime}{\partial y} \label{approx_qx_grho} \\
J_{q,y}^{approx} = -\left[\lambda + \Lambda_2 \right]  \frac{\partial T^{LS}}{\partial y} + d_y\frac{\partial T^\prime}{\partial y} 
\label{approx_qy_grho},
\end{align}
where we have introduced two new coefficients $d_x$ and $d_y$ to be fitted. 
The importance of using $\partial \rho / \partial y$ as a proxy for $\partial T^\prime / \partial y$ is highlighted in Appendix B, Fig. \ref{Measured_lambda_vs_startbin}, where different starting locations for the fitting profile can be seen to give very different measurement for heat flux. We have ignored nonlocal effects in the interface, which in principle should be included. However, the inclusion of nonlocal density coupling in shear flow is known to be a complex undertaking \citep{PhysRevE.92.012108, PhysRevE.91.062132} which is not warranted in this instance, and the approach used here provides a practical and convenient way to improve fitting in the presence of near-wall density stacking. 

The approximate expressions for fluxes, \eq{approx_qx_grho} and \eq{approx_qy_grho}, are fitted to the results for $J_{q,x}$, as shown in Fig. \ref{shear_heatflux_terms} (c) and $J_{q,y}$ in Fig. \ref{shear_heatflux_terms} (d),
with these fits giving values for $\Lambda_1$ and $\lambda + \Lambda_2$  respectively.
It seems reasonable to expect that we should be able get the coefficient of wall normal heat flux, $\Lambda_2$, by subtracting the coefficient of heat flux $\lambda$ obtained from \eq{sys1_lambda} at the measured densities and temperatures.
However, due to the highly coupled nature of the temperature and density highlighted in Fig \ref{TvsP}, as well as variation in the channel itself, this was found to provide poor results.
The values for $\Lambda_1$ and $\Lambda_2$ are found to be of similar magnitude to each other, $\sim 0.5$, but an order of magnitude smaller than the Fourier's law coefficient $\lambda$, which is of order $\sim 5$.
It is for this reason that these coefficients are difficult to obtain and temperature dependence becomes important. Further discussion of errors is provided in Appendix \ref{sec:error}.
However, as the two heat flux contributions $\Lambda_1$ and $\Lambda_2$ are known functions of strain rate, we can use this dependence as another route to the strain rate dependent thermal conductivity coefficients.


\begin{table}
\begin{center}
\begin{tabular}{|c|c|c|c|c|} \hline
$\rho_l$ & $T_l$ & System 2   & \eq{sys1_lambda}                & \eq{sys1_lambda}     \\
         &       & intercept  & $\lambda(\rho_l, T_l)$  & $\lambda(\rho_l, T_{wall})$      \\\hline
$\;0.65 \;$      & $\;0.9\;$    & 3.50                & 4.00                    & 3.37             \\\hline
$\;0.75\;$       & $\;0.95\;$   & 5.04                & 5.60                    & 4.89             \\\hline
$\;0.85\;$       & $\;1.0\;$    & 7.03                & 7.65                    & 6.77             \\\hline
\end{tabular}
\end{center}
\caption{Comparison of $\lambda$ predicted using parameterization obtained from results in system 1 at measured $T_l$ and $T_{wall}$, to the $\lambda$ value in system 2 using the zero strain rate intercept.}
\label{tab:lambda_compare}
\end{table}

To explore a range of strain rates, we compared coefficients from $L_y = 23.81$ to three extra domain sizes, $L_y = 47.5$, $L_y = 71.3$, $L_y = 95.2$, run with the same wall sliding velocity $v_w=\pm1$, each at three densities $\rho_l = 0.65$, $\rho_l = 0.75$ and $\rho_l = 0.85$.
In addition to varying system size, varying wall sliding velocities was also explored at $L_y = 23.81$ for $v_w = \{ 1.0, 2.0\}$ at densities $\rho_l = \{0.65, 0.75, 0.85\}$. 
However, for $v_w = 2.0$ we observe large pressure heating with temperature as high as $T\approx 4$ and, as a result, only varying system size was used in the strain rate analysis.

A plot of $\Lambda_1$ and $\Lambda_2$ against strain rate is presented in Fig. \ref{qx_qy2_vs_strain}. 
The parallel heat flux term $\Lambda_1=\lambda_1 \dot{\gamma}$ is linearly proportional to strain rate with a zero intercept, allowing us to use the fits of Fig. \ref{qx_qy2_vs_strain} to get a direct estimate of $\lambda_1$. 
The $\Lambda_2$ term is obtained by plotting the results for $\Lambda_2 + \lambda$ from the fits in Fig. \ref{shear_heatflux_terms} $d)$, against the strain rate and fitting to $\lambda + 3 \dot{\gamma}^2 \lambda_2$. The results are consistent with the expected quadratic strain rate dependence when $\lambda$ and $\lambda_2$ are allowed to vary freely in the fit. Subtracting the fitted intercept ensures a value of zero at the origin as required and gives us the results shown in Figure \ref{qx_qy2_vs_strain}.
The zero strain rate intercepts provide estimates of $\lambda$ for each density and the fit to the strain rate dependence provides a direct estimate of $\lambda_2$.
The fits to results for all the densities studied appear to be similar, suggesting that the $\lambda_2$ coefficient is not very sensitive to density within the range of densities studied here.

The values of $\lambda$ measured in system 2 are compared to the predictions using \eq{sys1_lambda} based on the liquid's density and temperature in Table \ref{tab:lambda_compare}. 
As the wall is thermostatted to $T_{wall}=0.7$, in the absence of shear the liquid would be expected to equilibrate to the same temperature as the walls.
The zero-strain intercept should therefore be the predicted $\lambda$ for this equilibrium case at $T_{wall}$, which is also shown in Table \ref{tab:lambda_compare}.
The system 2 prediction of $\lambda$ is seen to be close to the one expected at the wall temperature, within $5\%$, but with a slightly greater value in all cases.
The use of an average temperature over the liquid region or wall temperature is crude as both temperature and $\lambda$ would be expected to vary inside the channel.
A more detailed analysis could take into account the varying thermal conductivity coefficient as a function of $y$, using the approach of \citet{todd1997temperature}.

\begin{figure}
\includegraphics[width=0.48\textwidth]{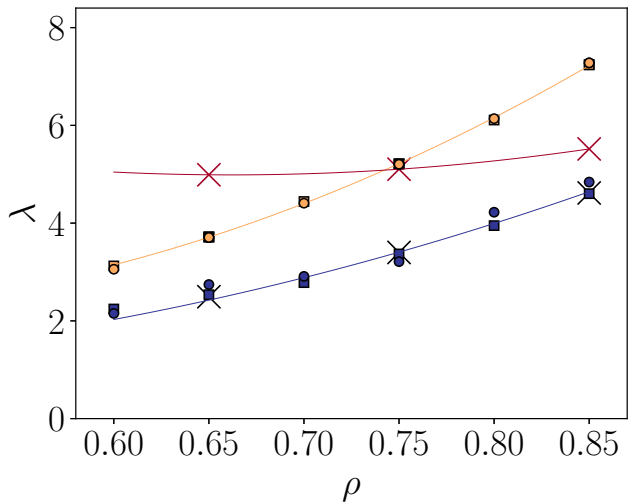}
\caption{Summary of coefficients from shear study with MOP shown by circles, VA by squares and crosses denoting the fits from Fig \ref{qx_qy2_vs_strain}. Blue is $\lambda_1$, gold $\lambda$ with $\lambda_2$ in red. Lines are included to guide the eye using a least squares quadratic polynomial fit to the data.}  
\label{shear_study_k}
\end{figure}

A summary of results for the three coefficients of thermal conductivity $\lambda$, $\lambda_1$ and $\lambda_2$ is presented in Fig. \ref{shear_study_k}. 
We use the strain rate $\dot{\gamma}^{LS} = \partial v^{LS}/\partial y$ to work out parallel heat thermal conductivity coefficients by simply dividing the VA and MOP values by the strain rate, $\lambda_1 = \Lambda_1/\dot{\gamma}^{LS}$ at each density.
As a cross check, we also use the fits obtained in Fig \ref{qx_qy2_vs_strain} over the range of strain rate studies, shown as crosses on Fig \ref{shear_study_k} with excellent agreement.
The Fourier's law thermal conductivity coefficients $\lambda$ are also presented in Fig. \ref{shear_study_k} for all studied density.
In each case, the value is corrected based on the average liquid temperature using \eq{sys1_lambda}.
The average liquid temperature measured in each system is as summarised in overview Fig. \ref{TvsP}.
Only $\lambda$ values for the highest strain rate case, $\dot{\gamma}=0.084$, are shown, other results are similar after correction.

Finally, the three $\lambda_2$ values obtained from the fits to strain rate of Fig \ref{qx_qy2_vs_strain} are also presented in Fig. \ref{shear_study_k}. A full discussion of possible sources of error is included in Appendix \ref{sec:error}.
Although the parallel heat flux has been previously observed in NEMD simulations of a dense fluid undergoing Poiseuille flow \cite{Han_Lee}, this is the first time that the values of both $\lambda_1$ and $\lambda_2$ have been evaluated directly from the heat fluxes. We further note that \citet{Han_Lee} observed a stronger heat flux in the parallel direction, in contrast to our much weaker heat flux observed for planar Couette flow. In future studies it will be interesting to repeat these measurements for Poiseuille flow to see whether we obtain similarly strong parallel heat fluxes, and measure the resulting values of $\lambda_1$ and $\lambda_2$.

The measured values of $\lambda_1$ in Fig \ref{shear_study_k} are found to be of similar magnitude to that found by \citet{daivis1993generalized} using a Green-Kubo like method, where a value of $\lambda_1 \approx 2.2$ was obtained for a LJ atomic system with a density of $\rho_l = 0.8$ and a temperature of 2.0. \citet{daivis1993generalized} also found that $\lambda_2$ was approximately zero, in contrast to the results presented here. The reason for this is not presently understood, but the different state point, LJ potential cutoff and range of shear rates studied may all be partially responsible.

Given the good agreement for $\lambda$ compared to liquid Argon experimental data, the presented measurements of $\lambda_1$ and $\lambda_2$ should be testable against laboratory experiments. 
The high shear rates required and small magnitudes of this secondary heat flux would make such measurements extremely challenging for this system. It is for this reason that molecular simulation remains an invaluable tool for studies of this kind.

\section{Conclusions}
\label{sec:Conclusions}

In this work we have used the \citet{Irving_Kirkwood} equations in integrated form to obtain expressions for heat flux, both as an average quantity in a volume (Volume Average, VA) and defined on surfaces (Method of Planes, MOP). 
By obtaining the heat flux on the various surfaces, the MOP approach is extended to provide heat flux parallel to the flow. 
The MOP and VA definitions are shown to give equivalent results for measured heat flux when properly averaged. 
From a non-equilibrium molecular dynamics (NEMD) simulation using a channel with a temperature gradient, the coefficient of heat flux from Fourier's law is obtained and shown to have good agreement to experimental results.
The Fourier's law coefficient is parameterised over a range of densities and temperatur es.
Boundary driven planar Couette flow is then used to explore the additional terms due to strain rate in the direction parallel and normal to the wall, as predicted by non-equilibrium thermodynamics. 

An estimate of the second coefficient of thermal conductivity in the wall normal direction is obtained by fitting a quadratic in strain rate.
This estimate of second coefficient is obtained at three different densities and appears to be broadly independent of density, within the range of densities studied in this work.
The zero strain intercept provides a separate estimate of the Fourier thermal conductivity coefficient, which is found to be slightly lower than the parameterised form obtained from system 1.
In addition, the results for the parallel component of heat flux appear to be very convincing.
As a function of strain rate, it is shown to vary linearly as predicted by theory, allowing us to fit over a wide range of strain rates providing this coefficient as a function of density.
The consistency of this coefficient from all simulations suggests this value could be used as a numerical prediction testable through experiments.

\appendix

\section{Additional Derivations}
\label{sec:A_derivation}
In the Appendix, Section \ref{sec:A_derivation} has a discussion of the alternative form of the \citet{Irving_Kirkwood} operator as an integral along a line, with the result that the velocity expressions can be defined along this line, as an average in a volume or as an average over a surface.
In Section \ref{sec:error}, the various sources of error in this work are set out and the attempts to mitigate them discussed, including the use of density gradients in the fitting.

The pressure from \citet{Irving_Kirkwood} is expressed as the difference between two Dirac delta functions. To derive an expression for the configurational part of the local, instantaneous pressure tensor, we will need to express the difference between two delta functions as a divergence. 
This can be done as follows. Using the fundamental theorem of calculus, we can write
\begin{align}
 \delta({\bf{r}} - {\bf{r}}_i) -  \delta({\bf{r}} - {\bf{r}}_j) = \int_0^1 \frac{\partial}{\partial s} \delta \left({\bf{r}} - \left[{{\bf{r}}_i} - s {\bf{r}}_{ij} \right] \right)  ds .
\end{align}
However,
\begin{align}
 \frac{\partial}{\partial s} =  \frac{\partial \left[{\bf{r}}_i - s {\bf{r}}_{ij} \right]}{\partial s}  \frac{\partial}{\partial \left[{\bf{r}}_i - s {\bf{r}}_{ij} \right]},
\end{align}
so we can write
\begin{eqnarray}
 \delta({\bf{r}} - {\bf{r}}_i) -  \delta({\bf{r}} - {\bf{r}}_j) = \;\;\;\;\;\;\;\;\;\;\;\;\;\;\;\;\;\;\;\;\;\;\;\;\;\;\;\;\;\;\;\;\;\;\;\;\;
 \nonumber \\
 \int_0^1 \frac{\partial}{\partial s} \delta \left({\bf{r}} - \left[{\bf{r}}_i - s {\bf{r}}_{ij} \right] \right)  ds
 \nonumber \\
= {\bf{r}}_{ij} \cdot \int_0^1  \frac{\partial}{\partial \left[{\bf{r}}_i - s {\bf{r}}_{ij} \right]} \delta \left({\bf{r}} - \left[{\bf{r}}_i - s {\bf{r}}_{ij} \right] \right)  ds
 \nonumber \\
= \frac{\partial}{\partial {\bf{r}}} \cdot {\bf{r}}_{ij}  \int_0^1  \delta \left({\bf{r}} - \left[{\bf{r}}_i - s {\bf{r}}_{ij} \right] \right)  ds,
\end{eqnarray}
which is the required form as an integral along the line between the atoms.
The volume integral of this expression is as given in \eq{int_s}, which occurs in quantities which depend on inter-particle interactions such as configurational pressure and heat flux.
The heat flux includes velocity and it is necessary to write the averaged fluid velocity integrated along the line,
\begin{align}
{\mathbf{\bar v}_s}\left( {{{\mathbf{r}}_{ij}}}, {{{\mathbf{r}}_{m}}} \right) = \;\;\;\;\;\;\;\;\;\;\;\;\;\;\;\;\;\;\;\;\;\;\;\;
\nonumber \\
 {{\int_0^1 {{\mathbf{v}}\left( {{{\mathbf{r}}_s }} \right){\vartheta _s }ds } } \mathord{\left/
 {\vphantom {{\int_0^1 {{\mathbf{v}}\left( {{{\mathbf{r}}_s }} \right){\vartheta _s }ds } } {\int_0^1 {{\vartheta _s }ds } }}} \right.
 \kern-\nulldelimiterspace} {\int_0^1 {{\vartheta _s }ds } }} = 
 \nonumber \\
{{\int_0^1 {{\mathbf{v}}\left( {{{\mathbf{r}}_s }} \right){\vartheta _s }ds } } \mathord{\left/
 {\vphantom {{\int_0^1 {{\mathbf{v}}\left( {{{\mathbf{r}}_s }} \right){\vartheta _s }ds } } {{\ell _{ij}}}}} \right.
 \kern-\nulldelimiterspace} {{\ell _{ij}}}}.
\label{v_lambda}
\end{align}
Then the full expression for the volume averaged heat flux vector is
\begin{align}
\heatflux{}_V = \frac{1} {\Delta V} \int_{\intV}  \heatflux \left( \textbf{r},t \right) \dV \;\;\;\;\;\;\;\;\;\;\;\;\;\;\;\;\;\;\;\;\;\;\;\;\;\;\;\;\;\;\;\;\;\;\;\;\;\;\;\;\;\;\;\;\;\;\;\;
\nonumber \\
= \frac{1} {\Delta V} \left[ \displaystyle\sum_{i=1}^N  \MDpenergy \MDpvel \vartheta_i -  \frac{1}{2} \displaystyle\sum_{i,j}^N \Fijrij \cdot \left( \MDvel   - {\mathbf{\bar v}_s}\left( {{{\mathbf{r}}_{ij}}}, {{{\mathbf{r}}_{m}}} \right)  \right) \ell _{ij} \right]  .
\label{VA_heatflux}
\end{align}
Note this includes a contribution from the difference between the velocity $\MDvel$ of particle $i$ and the fluid velocity averaged over all points on the inter-molecular interaction line $\mathbf{r}_{ij}$ that are inside $\Delta V$, ${\mathbf{\bar v}}\left( {{{\mathbf{r}}_{ij}}} \right)$.  If $v$ is constant inside $\Delta V$, then the $\ell_{ij}$ on the top and bottom cancel and we are left with $\bf{v}$ at ${\bf{r}}_m$ as assumed in equations (\ref{EnergyEqminusadvct2_VA}) and (\ref{VA_stressheat}).

In order to get the expressions for the heat flux, we can convert the kinetic part of the heat flux vector in \eq{EnergyEqminusadvct2_VA} into a form where the instantaneous streaming velocity is not required. We can write
\begin{align}
  {\mathbf{J}}_q^K\left( {{\mathbf{r}},t} \right) &= \sum\limits_i^N {{u_i}{{\mathbf{c}}_i}\delta \left( {{\mathbf{r}} - {{\mathbf{r}}_i}} \right)}  \nonumber \\ 
   &= \sum\limits_i^N {\left( {{e_i} - \frac{1}{2}{m_i}{{\mathbf{v}}^2}\left( {\mathbf{r}} \right)} \right)\left( {{{\mathbf{v}}_i} - {\mathbf{v}}\left( {\mathbf{r}} \right)} \right)\delta \left( {{\mathbf{r}} - {{\mathbf{r}}_i}} \right)}  \nonumber \\ 
   &= \sum\limits_i^N {{e_i}{{\mathbf{v}}_i}\delta \left( {{\mathbf{r}} - {{\mathbf{r}}_i}} \right)}  - {\mathbf{v}}\left( {\mathbf{r}} \right)\sum\limits_i^N {{e_i}\delta \left( {{\mathbf{r}} - {{\mathbf{r}}_i}} \right)}
\end{align}
where the last line follows from
\begin{align}
\sum\limits_i^N {{m_i}\left( {{{\mathbf{v}}_i} - {\mathbf{v}}\left( {\mathbf{r}} \right)} \right)\delta \left( {{\mathbf{r}} - {{\mathbf{r}}_i}} \right)}  
\nonumber \\
= \sum\limits_i^N {{m_i}{{\mathbf{v}}_i}\delta \left( {{\mathbf{r}} - {{\mathbf{r}}_i}} \right)}  - {\mathbf{v}}\left( {\mathbf{r}} \right)\sum\limits_i^N {{m_i}\delta \left( {{\mathbf{r}} - {{\mathbf{r}}_i}} \right)} = 0.
\end{align}

In the main text, we use the velocity averaged along a line and in a volume when we consider heat flux in a volume; while for the surface or plane based measures, velocity is defined on the surface as follows,
\begin{align}
\int_{S_x^+} \CFDvel ({\bf{r}}, t)\cdot  d{\textbf{S}_x^+}  \define \;\;\;\;\;\;\;\;\;\;\;\;\;\;\;\;\;\;\;\;\;\;\;\;\;\;\;\;\;\;\;\;\;\;\;\;\;\;\;\;\;\;\;\;\;\;\;\;\;\;\;\;\;\;\;\;\;\;\;\;\;\;\;
\nonumber \\
 \frac{1}{\sum_{i=1}^N  m_i \vartheta_i}\sum_{i=1}^N  \MDvel \delta \left( {{x_i} - {x_ + }} \right)\Pi \left( {\frac{{{y_i} - {y_m}}}{{\Delta y}}} \right)\Pi \left( {\frac{{{z_i} - {z_m}}}{{\Delta z}}} \right),
\label{vel_definition_MOP}
\end{align}
where this is the velocity on the a plane located at $x^+$ obtained from the average flux of atoms \citep{Todd_et_al_95} and the density, obtained as an average over a cell centred on $\bf{r}$, as $ \sum m_i \vartheta_i$ should be extrapolated to the location of the cell surface.
An alternative approach would be to collect density at the location of the surface,
\begin{align}
\int_S \CFDvel ({\bf{r}}, t)\cdot  d{\textbf{S}_x^+}  \define \;\;\;\;\;\;\;\;\;\;\;\;\;\;\;\;\;\;\;\;\;\;\;\;\;\;\;\;\;\;\;\;\;\;\;\;\;\;\;\;\;\;\;\;\;\;\;\;
\nonumber \\
 \frac{\sum_{i=1}^N m_i \MDvel  \delta \left( {{x_i} - {x_ + }} \right)\Pi \left( {\frac{{{y_i} - {y_m}}}{{\Delta y}}} \right)\Pi \left( {\frac{{{z_i} - {z_m}}}{{\Delta z}}} \right) }{\sum_{i=1}^N  m_i  \delta \left( {{x_i} - {x_ + }} \right)\Pi \left( {\frac{{{y_i} - {y_m}}}{{\Delta y}}} \right)\Pi \left( {\frac{{{z_i} - {z_m}}}{{\Delta z}}} \right)},
\label{vel_definition_MOP_better}
\end{align}
where the density at a plane can be computed using the method of \citet{Daivis_et_al}.
In practice, both approaches are used and compared, giving similar results in this work as the statistics collected are in steady-state temperature and shear driven flows.
However, in more general cases such as turbulent flow \citep{Smith_2015} this distinction may become important.

\section{Uncertainty Analysis}
\label{sec:error}

We discuss the sources of possible error in this study, which include: (i) temperature variation, (ii) finite system size effects, (iii) departure from classical Navier-Stokes hydrodynamics and (iv) statistical errors.

First, consider (i) temperature due to shear heating and the thermostatting of only the wall atoms. Different density fluids have different average temperatures, with denser fluids generally hotter. In the worst case of $\rho_l=0.85$ and a channel of height $L_y=23.81$ in system 2, the mean fluid temperature increases to $T=1.25$, which is compared to the value of $T=1.0$ at the same density when obtaining $\lambda$ in system 1. The difference in heat flux due to the difference in temperature, estimated using NIST isochoric data at $\rho_l=0.85$ for liquid Argon \citep{Perkins1991}, would give a difference of $\sim 2.5\%$ in the value of $\lambda$, which could imply an almost $60\%$ error in the calculation of $\lambda_2$. This is a direct consequence of the relative magnitudes of the $\lambda$ and $\lambda_2$ coefficients, making isolation of these higher order heat flux terms difficult. It is also unclear if $\lambda_2$ would be expected to have the same functional dependence on temperature as $\lambda$.
It should be noted that strain rate dependence is very difficult to decouple from temperature dependence without thermostatting the entire system; an approach which should be avoided as previous work has demonstrated that wall-only thermostatting is essential in order is preserve the correct system dynamics \citep{Bernardi10a, Bernardi10b}.
The use of temperature correction based on system 1 has been employed in this work in an attempt to solve this dependence.

Another possible source of error, (ii), is finite size effects. In the work of \citet{Hyzorek_Tretiakov}, the effect of channel height on measured heat flux shows that confinement effects cause a departure from the experimental Fourier's law. This is observed for channels smaller than about $20$ in reduced units, especially at higher densities. Although \citet{Hyzorek_Tretiakov} use a longer cutoff radius $r_c = 5$ and an infinite wall-fluid interaction length (factors which potentially increase the impact of nano-confinement) these trends suggest that the value of $\lambda$ could be reduced by the smaller channel used in system 2 as compared to system 1. This would, in turn, suggest that the obtained $\lambda_2$ in this work may increase in error for high densities. The departures observed by \citet{Hyzorek_Tretiakov} are of the order $1\%$ when applied to get $\lambda_2$ for low density systems ($\rho_l=0.4$), increasing to $10\%$ for the highest densities they considered ($\rho_l=0.7$). 
Extrapolation using a fit to digitised data from \citet{Hyzorek_Tretiakov} suggest this may rise as high as $25\%$ in the highest density, $\rho_l = 0.85$, presented in this work. Again, it is not clear if the $\lambda_1$ and $\lambda_2$ coefficients will be similarly impacted by nano-confinement.
The use of larger systems, as large as $L_y = 95.2$ reduced units in the strain rate study, help reduce this effect.
However, large channels necessarily have low strain rates and so poor signal to noise ratio. Fitting over a range of strain rates is used to get values of $\lambda_1$ and $\lambda_2$.

The uncertainty induced due to (iii) departures from classical hydrodynamics include density oscillations, non-zero slip lengths, non-Newtonian effects and departure from expected velocity and temperature profiles. Indeed, a departure from a linear velocity profile is observed in the high density large channel cases (greater than $40$ reduced units), an effect especially apparent near the wall. As a result, the measured strain rate is taken not simply as $v_w/L_y$ but obtained as the average profile of velocity over the domain. This is the reason that the strain rates in Fig. \ref{qx_qy2_vs_strain} are different for the three different densities. No wetting parameter is used in this work but the higher density walls will potentially cause slip and Kapitza-like jumps \citep{HungKim_2008}. The impact of slip-length, both in temperature and velocity profiles, has been reduced by fitting only to the liquid region, \ie{} applying fitting to averaging bins located a distance from the wall. Slip is linked to 
density stacking effects discussed previously and so the inclusion of density in our fits also help to reduce this effect, as discussed in point (iv).

\begin{figure}
    \begin{subfigure}[t]{0.45\textwidth}
    \includegraphics[width=\textwidth]{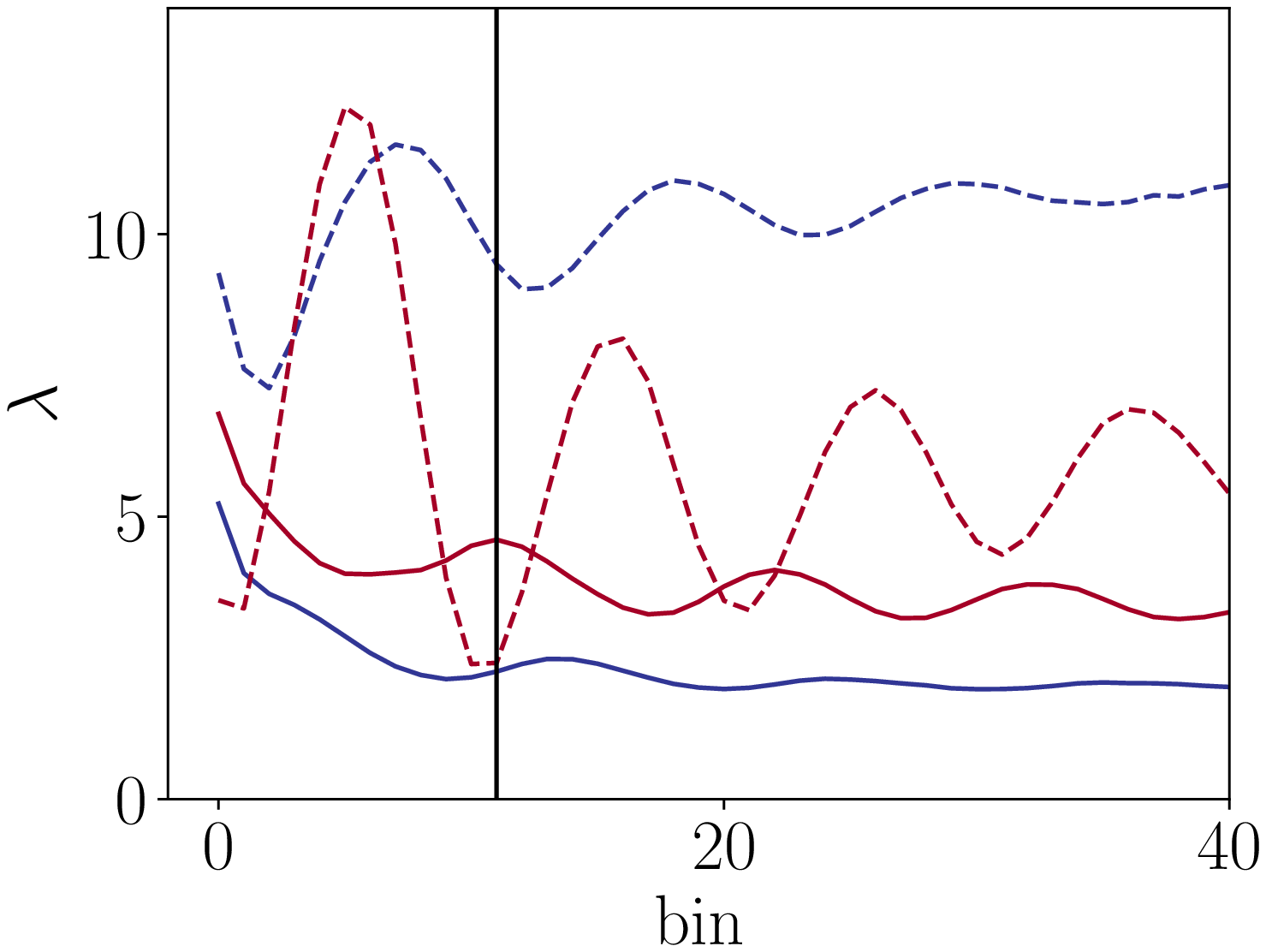}
    \put(-175,145){$a)$}
  \end{subfigure} 
  \begin{subfigure}[t]{0.45\textwidth}
    \includegraphics[width=\textwidth]{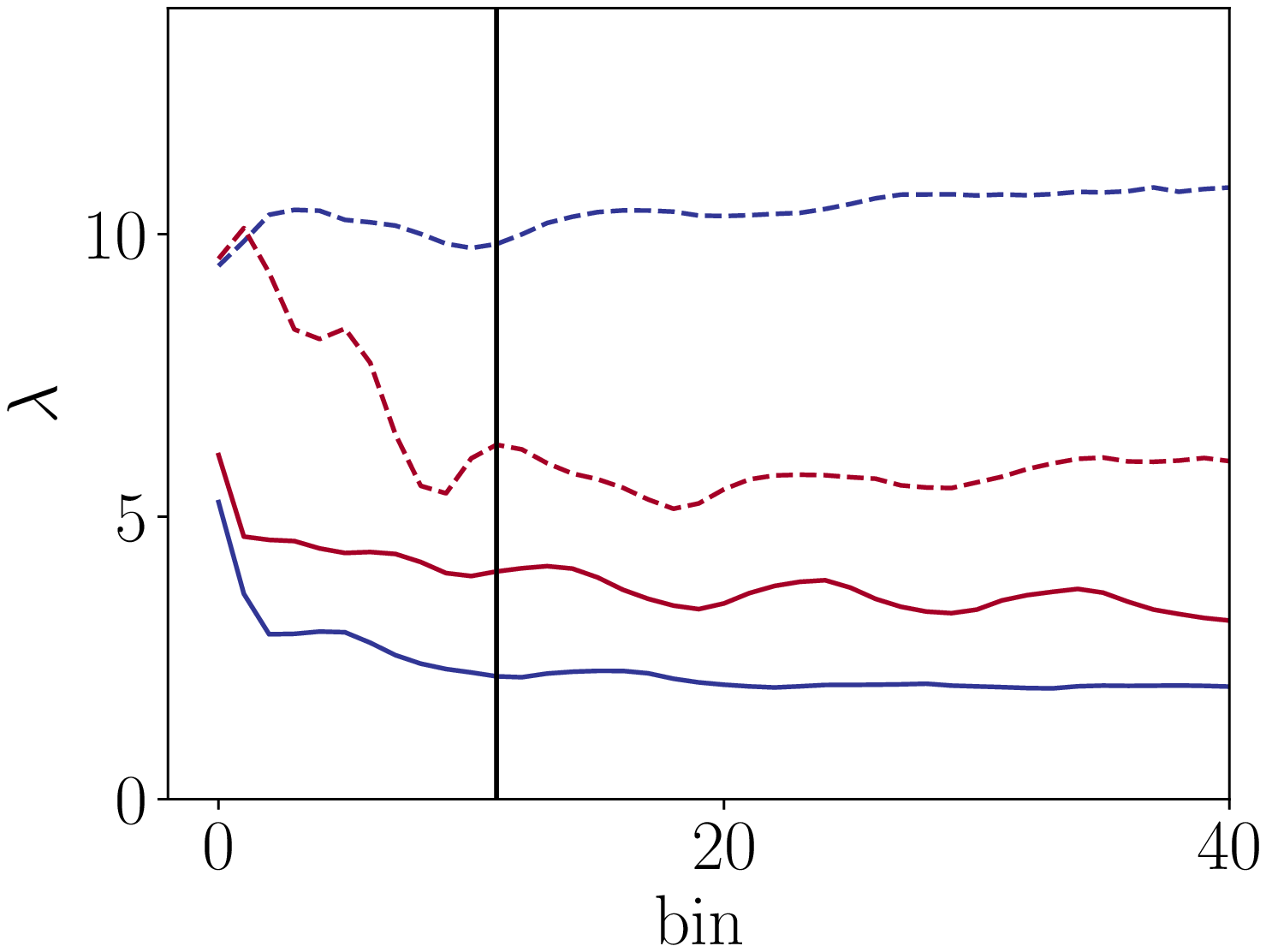}
    \put(-175,145){$b)$}
  \end{subfigure} 
  \caption{The starting bin for fit (relative to the wall) on the measured VA heat flux coefficients $\lambda_x$ (solid lines) and $\lambda_2$ (dotted lines) with two densities $\rho_l = 0.6$ in blue and $\rho_l = 0.8$ in red. $a)$ shows before including the $\partial \rho / \partial y$ correction while $b)$ includes this term. A bin value of 11 is chosen as the starting point for fits in all results presented in this work, shown by the black line on the figure.}
  \label{Measured_lambda_vs_startbin}
\end{figure}

These measurements are also subject to (iv) large levels of noise with very weak signals, shown for example in Fig. \ref{diff_shear_energy_terms} (a). 
These errors have been reduced as far as possible by collecting statistics for large atomistic systems over long runs.
In addition, the use of least square fits to obtain derivatives of temperature and velocity improve statistics.
However, the density stacking near the wall poses a problem for fitting least squares solutions to obtain coefficients of heat flux.
To highlight the problem, consider Fig \ref{Measured_lambda_vs_startbin} $a)$ which shows the fitting parameter (\ie{} gradient of the line) returned when we start the fit at different bins in the channel.

\begin{figure}
\includegraphics[width=0.48\textwidth]{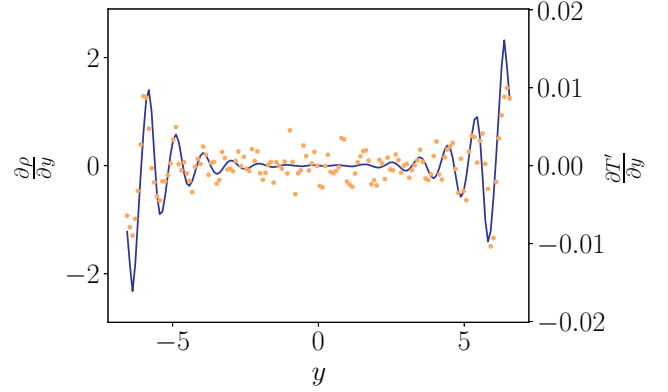}
\caption{Comparison of the shape of density component of temperature gradient $\partial T^\prime / \partial y$ (gold points) and the density gradient (blue line) which is used to replace them, including the density stacking but reducing the measured noise in the gradient.}  
\label{gradient_density_and_temperature}
\end{figure}

In this case, a value of zero on the $x$-axis means a fit to all $170$ liquid bins, while $10$ for example excludes the $10$ bins either side of the channel closest to the wall. This means we would fit a line to only $150$ innermost ones.
By including the part of the channel with density peaks, it is clear that we obtain completely different coefficients, by almost an order of magnitude in Fig \ref{Measured_lambda_vs_startbin} $a)$.
This is a particular problem in this work given the very small magnitudes of the $\lambda_1$ and $\lambda_2$ coefficients compared to variations in density in this highly confined channel.
As a result, it is almost impossible to work out a meaningful location to start fitting and using only the inner-most part of the channel gives very poor statistics.
Instead, we improve the fit by replacing the term for the oscillating component of the temperature field $\partial T^\prime / \partial y$ with a term proportional to the gradient of density itself $\partial \rho / \partial y$, as the gradient of density correlates very well with the gradient of temperature but has much less noise as shown in Fig \ref{gradient_density_and_temperature}. \citet{todd1997temperature} have shown that this is a good approximation near the centre of the channel when the pressure is constant, which is what we can expect when local thermodynamic equilibrium holds. In the interfacial region, the thermodynamics is more complex but the density and temperature gradients remain highly correlated.
By including the density derivative term in \eq{approx_qx_grho} and \eq{approx_qy_grho}, the results as a function of fitting location become much less susceptible to density peaks, as shown in Fig \ref{Measured_lambda_vs_startbin} $b)$.
The starting point for the fitting is chosen to be bin $11$, shown by the black line on Fig \ref{Measured_lambda_vs_startbin} $b)$ as this appears to be consistent with the average gradient in the channel and includes the majority of bins in the fit. 

\end{document}